\documentstyle[12pt,aaspp4]{article}

\lefthead{Steindling et al.}
\righthead{Str\"{o}mgren photometry from $z=0$ to $z \approx 1$}

\begin{document}
\title{Str\"{o}mgren photometry from $z=0$ to $z \approx 1$. I. The method}
\author{Susanna Steindling\altaffilmark{1}, Noah Brosch\altaffilmark{2}}
\affil{Wise Observatory, School of Physics and Astronomy \\ Tel Aviv University, 
Tel Aviv
69978,  
Israel}
\and
\author{Karl D. Rakos\altaffilmark{3}}
\affil{Institute for Astronomy, University of Vienna, T\"{u}rkenschanzstra\ss e 
17, 1180 Vienna, Austria}
\altaffiltext{1}{susan@wise.tau.ac.il}
\altaffiltext{2}{noah@wise.tau.ac.il}
\altaffiltext{3}{rakos@astro.univie.ac.at}


\begin{abstract}
We use rest-frame Str\"{o}mgren photometry to observe clusters of galaxies in a 
self-consistent manner from  {\em z}\/=0 to {\em z}\/=0.8. Str\"{o}mgren 
photometry of galaxies is intended as a compromise between standard broad-band 
photometry and spectroscopy, in the sense that it is more sensitive to subtle 
variations in spectral energy distributions than the former, yet much less 
time-consuming than the latter. Principal Component Analysis (PCA) is used to 
facilitate extraction of information from the Str\"{o}mgren data. By calibrating 
the Principal Components using well-studied galaxies, as well as models of 
stellar populations, we develop a purely empirical method to detect, and 
subsequently classify, cluster galaxies at all redshifts smaller than 0.8. 
Interlopers are discarded with unprecedented efficiency (up to 100 \%). The 
first Principal Component essentially reproduces the Hubble Sequence, and can 
thus be used to determine the global star formation history of cluster members. 
The (PC2, PC3) plane allows us to identify Seyfert galaxies (and distinguish 
them from starbursts) based on photometric colors alone. In the case of E/S0 
galaxies with known redshift, we are able to resolve the age-dust-metallicity 
degeneracy, albeit at the accuracy limit of our present observations.
We use this technique in later papers to probe galaxy clusters well beyond their 
cores and  to fainter magnitudes than spectroscopy can achieve, because the 
faint end of the luminosity function as well as the outer cluster regions seem 
to exhibit the strongest evolutionary trends.  We are able to directly compare 
these data over the entire redshift range without a priori assumptions because 
our observations do not require first-order k-corrections. The compilation of 
such data for different cluster types over a wide redshift range is likely to 
set important constraints on the evolution of galaxies and on the clustering 
process.
\end{abstract}
\keywords{galaxies: clusters: general --- galaxies: evolution --- galaxies: 
fundamental parameters --- galaxies: photometry --- methods: data analysis --- 
techniques: photometric}

\section{Introduction}
The study of galaxy evolution has benefited greatly from observations of galaxy 
clusters, because of better statistics (compared to pencil beam surveys), and 
because clusters are more easily identified  than isolated galaxies at large 
redshifts (e.g., Couch et al. 1983, Dressler 1993). These studies revealed some 
of the most spectacular galaxy evolutionary trends, such as the
Butcher-Oemler effect (Butcher \& Oemler 1978). Since then, clusters of galaxies 
have been used as benchmarks to sample galaxy evolution at different redshifts 
(e.g., Butcher, Oemler \& Wells 1983, Dressler et al. 1999).

However, data on galaxies in clusters are usually limited to the inner regions 
and/or to the brighter cluster members because of technical constraints; this 
selection effect biases the selection and comparison of clusters at different 
redshifts, as the fields of view and apparent magnitudes translate into 
different metric sizes and luminosities. The observed changes may therefore not 
be straightforward to interpret in terms of galactic evolution (e.g., Andreon \& 
Ettori 1999). Moreover, the galaxies are probably affected by the evolution of 
clustering itself. The modern view of hierarchical clustering (e.g., Baugh, Cole 
\& Frenk 1996) suggests that the latter might outweigh the former. Thus, many 
degeneracies, both intrinsic and technical, plague the study of distant cluster 
galaxies. By intrinsic, we mean the difficulty to decide whether the observed 
redshift evolution of the galaxies is due to cosmological galaxy evolution, or 
rather the result of environmental effects, i.e., changes in cluster properties 
that affect the galaxies in it. Resolving this issue requires a coverage of the 
parameter space (redshift, cluster type, radial dependence,...) that does  not 
yet exist.

 The method described below is aimed at bringing this kind and quantity of data  
within reach. It is an observing and analysis technique that enables one to 
probe the luminosity function to fainter objects than can be reached 
spectroscopically, as well as to cover more sparsely populated regions, such as 
poor clusters and the outskirts of rich ones, in a {\em consistent} manner for 
all redshifts up to {\em z}\/=0.8 (instrumental limitation). It conserves the 
advantages of classical photometry (depth and spatial coverage) but avoids its 
usual pitfalls, such as field contamination, k-corrections and loss of essential 
spectral information. 

This paper is organized as follows: section 2 outlines the advantages of our 
rest-frame photometric observing technique. In the following sections we 
describe the properties of such data using Principal Component Analysis: in 
section 3 we show how to define cluster members in PC-space (the 
``cluster-box''), and demonstrate how efficiently stars as well as fore- and 
background galaxies are recognized in this new parameter space. After 
eliminating the interlopers, we show, in section 4, how to differentiate the 
various  types of cluster members in that same parameter space, which is 
necessary to address questions of galaxy evolution. Effects of age, dust and 
metallicity, as well as AGN activity are discussed. Section 5 summarizes the 
method.

\section{A new approach to extra-galactic photometry}
\subsection{Limitations of standard broad-band photometry}
A spatially resolved galaxy spectrum is the richest information harvest an 
observer can hope for. However, at  $z \gtrsim 0.3$ galaxies are barely a few 
arcseconds in size, and one has to settle, in most cases, for a single aperture 
spectrum.
Ideally, one wishes to obtain the spectra of all the galaxies in a given 
cluster, thereby determining their redshift and cluster membership, and infer 
their star formation histories by comparing these spectra to evolutionary 
models. Unfortunately, even with the largest telescopes, this requires 
unreasonably large amounts of observing time, if one wishes to cover the entire 
range of cluster types and redshifts. Moreover, multi-slit or fiber spectroscopy 
introduces a selection bias by rendering simultaneous observations of close 
pairs or subgroups technically difficult and time-consuming, because of 
adjacency limitations due to the physical size of the fibers. Aperture effects, 
related to the fraction of the galaxy being sampled at different distances, slit 
position, etc., constitute additional complications.

The usual alternative is broad-band photometry, with which the pioneering work 
on distant galaxy clusters has begun. 
However, the gain in time offered by this method is often outweighed by the loss 
of spectral information. Firstly, most existing studies deal with field 
contamination only statistically (e.g., Wilson et al. 1997). Although 
photometric redshifts are in principle of good quality (see e.g., Connolly et 
al. 1995), they rely heavily on the U-band to catch the 4000 \AA \, break at $z 
\leq 0.4$ or the rise in UV flux of star forming galaxies for $z \leq 0.6$, but 
the U-band is usually not used in high-z cluster observations aimed at the 
visible range of the cluster galaxie's SED.

Secondly, the large width of standard broadband filters causes  smearing of 
spectral information. Hence, comparing  photometric data to evolutionary 
synthesis models suffers from various degeneracies, such as the age-metallicity 
degeneracy (Worthey 1994).

An additional major problem arises when comparing photometric data from clusters 
at different redshifts, because the standard filters sample completely different 
regions of the rest-frame spectra.
Such comparisons require large k-corrections, which are extremely 
morphology-dependent and presumably also evolution-dependent. In fact, they 
require {\it a priori} knowledge of the evolutionary effects one is looking for, 
a circular argument. In some lucky cases, standard filters at one redshift 
correspond roughly to other standard filters at another redshift (e.g., 
Stanford, Eisenhardt \& Dickinson 1995, 1997), yet differential k-corrections 
still have to be made, and the datasets thus generated are somewhat 
heterogeneous and of limited common spectral coverage.

Lastly, the comparison of galaxies at various redshifts in search for 
evolutionary  effects requires a meaningful selection criterion, which is valid 
over the entire z-range. This, too, is plagued by the inherent uncertainty of 
the k-corrections, which differ by up to one magnitude for different Hubble 
types (de Vaucouleurs, de Vaucouleurs \& Corwin 1976). 

\subsection{The Str\"{o}mgren Photometry} 
Rakos and coworkers have pioneered an observing technique that resolves, or at 
least alleviates, some of the aforementioned problems: extra-galactic  {\em 
rest-frame} Str\"{o}mgren photometry (see Rakos \& Schombert 1995 and references 
therein). Unlike the Johnson system, the Str\"{o}mgren filters have been 
intentionally {\it designed to match} specific signatures in the spectra of 
stars, that relate directly to the  physical properties one wishes to 
investigate, such as temperature, metallicity, and surface gravity 
(Str\"{o}mgren 1966). 
For technical reasons, Rakos et al. slightly modified the bandpass definitions, 
so that all four filters are now ${\rm \sim 200 \AA }$  wide, and their central 
wavelengths are {\em uz}\/=3500~\AA, {\em vz}\/=4100~\AA, {\em bz}\/=4675~\AA, 
and {\em yz}\/=5500~\AA \/ respectively. (The lower case $z$ in the filter name 
refers to the rest-frame of the source.) These slight modifications do not 
influence the interpretation of the photometry, therefore we will not 
systematically distinguish ``Str\"{o}mgren filters'' from ``modified 
Str\"{o}mgren filters'' hereafter. We will omit the {\em z} and refer to these 
bands as  $(u, \, v, \, b, \, y)$, understanding implicitly that they are 
``tuned'' to the rest-frame of the target cluster.

Rakos, Schombert \& Kreidl (1991), Rakos \& Schombert (1995), and Rakos, Maindl 
\& Schombert (1996), describe the original stellar interpretation of 
Str\"{o}mgren fluxes and colors, and show how the same quantities can be used to 
characterize extra-galactic objects. Briefly, for homogeneous stellar 
populations, the  $(u\!-\!v)$ color measures the strength of the 4000\AA \, 
break, which can be used as an indicator of recent star formation.  The $(b\! 
-\! y)$ color measures the slope of the continuum redward of the break. The $b$ 
and $y$ filters are situated in regions free of any prominent absorption 
features, thus $(b\!-\!y)$ should be a good indicator of mean stellar age, free 
of metallicity effects. The $v$ filter, on the other hand, contains the region 
of the FeI+CN line blend. The photometric index $m\!=\!(v\!-\!b)\!-\!(b\!-\!y)$ 
can therefore be used to measure metallicity effects. Needless to say, when 
measuring real galaxy SEDs the colors are less straightforward to interpret, as 
the effects of mixed stellar populations and internal dust extinction are 
difficult to account for. For example, the ${\rm H_{\delta}}$ line present in 
the spectra of hot stars introduces an age-dependence in the $v$ filter. 
Therefore, we will rely in what follows on a more empirical approach.

In summary, the main advantages of this observing technique are: (1) the filters 
are designed to sample spectral regions, which are very sensitive to changes in 
the underlying physical properties, (2) these filters avoid all strong emission 
lines, which cause confusion in standard UVB photometry of active galaxies, and  
(3) observations are carried out at {\em fixed rest-frame} wavelengths, thus 
avoiding the uncertain k-corrections and delivering a self-consistent data set 
over the entire redshift range covered. Of course, this requires a priori 
knowledge of the cluster's redshift. 

With optical telescopes+cameras, the rest-frame Str\"{o}mgren method can be 
applied to clusters from $z=0$ to $z \sim 1$. However, the redshift segments 
$0.36 < z < 0.41$, $0.60 < z < 0.66$ and $0.83 < z < 0.89$ need special 
attention because of the atmospheric A-band contaminating the redshifted $y$, 
$b$ or $v$ filter respectively. This can in principle be calibrated out with an 
adequate spectrophotometric standard star, otherwise, these regions should be 
avoided.

\section{Identification of cluster members via PCA}
The basic approach adopted here is to assume that cluster galaxies concentrate 
in a specific location of the three-dimensional space defined by the 
Str\"{o}mgren colors. We demonstrate this below for a number of field and 
cluster galaxy samples.

\subsection{ From colors to PCA space: Definition of the ``cluster-box''}
We collected large aperture spectra from the literature (Kennicutt 1992 and 
Kinney et al. 1996) and convolved them with synthetic Str\"{o}mgren filter 
response curves, to simulate the appearance of known galaxy types in the 
three-dimensional $[(u\!-\!v), \, (v\!-\!b), \, (b\!-\!y)]$ color space 
(Fig.~1). The details of the samples are given in Tables 1 and 2. We insist on 
the ``large aperture'' selection criterion  (although it limits the statistics 
of the template sample) because we will subsequently compare them to  high 
redshift aperture photometry, and we want to avoid aperture-related color 
effects. Indeed, the color difference due to varying aperture sizes can be 
larger than the intrinsic color difference of different galaxy types observed 
through a constant aperture (see e.g., de Vaucouleurs, de Vaucouleurs \& Corwin 
1976 and Brosch \& Shaviv 1982). This effect is expected to be even more severe 
in Str\"{o}mgren colors, which are more sensitive to differences in stellar 
population (and their gradients). As will be shown at the end of this section, 
the sample nevertheless covers well the entire range of nearby galaxy types.

It is clear from Figure 1 that the ensemble of galaxies occupies a well 
confined, but difficult to fathom  subspace, which we now want to define in the 
simplest possible way. Since the default coordinate system (consisting of the 
three axes $ u\!-\!v$, $ v\!-\!b$ and $b\!-\!y$) is impractical for this 
purpose, we performed a Principal Component Analysis (PCA) on the data points. 

Briefly, a Principal Component Analysis, in any n-dimensional space, calculates
the axis along which the data points present the largest, most significant 
scatter. This is called the first Principal Component (PC1). It then proceeds to 
calculate  PC2, the axis of the second most significant spread in the remaining
n-1 dimensional space orthogonal to the first Principal Component, and so on.
Mathematically, this is done by  normalizing the coordinates of the data points 
to standardized variables and calculating their covariance. The final output  
are the eigenvectors (the Principal Components) and eigenvalues of this 
covariance matrix.
If, at one point, the standard deviation of the m-th Principal Component is no 
larger than the accuracy of the data, it means that the data can be fully 
described by only n-m components. The gain of the PCA is therefore twofold: (1) 
it minimizes the dimensionality of the data and (2) it provides  an orthonormal 
coordinate system in which the data are most easily characterized. The reader is 
referred to Lahav et al. (1996) for a more detailed description of PCA in an 
astronomical context.
 
\newpage
The base vectors of the new coordinate system, in which we will from now on view 
the data, are :
\begin{eqnarray*}
PC1 & = & 0.80 (u-v) + 0.53 (v-b) + 0.28 (b-y) \\
    & = & 0.8 u - 0.27 v - 0.25 b - 0.28 y \\
PC2 & = & -0.56 (u-v) + 0.49 (v-b) + 0.67 (b-y) \\
    & = & -0.56 u + 1.05 v + 0.18 b - 0.67 y \\
PC3 & = & 0.22  (u-v) - 0.69 (v-b) + 0.69 (b-y) \\
    & = & 0.22 u - 0.91 v + 1.38 b - 0.69 y
\end{eqnarray*}
Figure 2 gives a view of Kennicutt's and Kinney's galaxies in this new 
coordinate system (PC-space). Note that the only difference between Figures 1 
and 2 (i.e., the transformation matrix between color-space and PCA-space) is a 
simple 3D rotation. Yet, Figure 2 appears much simpler! The PC-space not only 
renders features more easily identifiable to the eye, it also facilitates 
automatic machine treatment of the data by providing a set of independent 
variables.
The new coordinates being a linear combination of the original ones, we refer to 
them as ``colors'' too in what follows. Note that PC2 and PC3 can also be 
understood as curvature indices, as in Koo (1985).

Since galaxy spectra vary continuously along (and across) the Hubble sequence 
(Kennicutt 1992a, 1992b), the color space they occupy must also be continuous. 
We therefore choose to define the allowed space for  cluster galaxies as the box 
defined by the maximal extent of the distribution in the new coordinate space 
(see Fig. 2). Because our observations are made at fixed {\em rest-frame} 
wavelengths, this cluster-box is essentially invariant with redshift. One might 
argue that differential k-corrections will still have to be made, because the 
rest-frame {\em width} of our filters changes with redshift, but these 
corrections (1) are very small, and (2) can easily be avoided by redshifting the 
template spectra and recalculating their colors. This will only be required if 
high precision photometry ($dm \sim 0.01 - 0.02$ ) is available or for 
comparisons of clusters over $dz > 0.4 $. If one wants to avoid this additional 
computation step, it is permissible to simply use the values obtained for 
templates shifted to $z=0.4$ (the middle of our redshift range). As can be seen 
in Figure 3, the differences in PC-colors due to the spectral stretching are 
negligible in most cases and can be accounted for by a slight widening of the 
cluster-box boundaries when the difference for certain spectral types becomes of 
the order of the PC-color accuracy. 
\newpage
In  PC-space, the averages and standard deviations of the galaxies' coordinates 
are:

\[<PC1>=0.35  \ \ \sigma_{PC1}=0.45 \]
\[<PC2>=-0.18 \ \  \sigma_{PC2}=0.10 \]
\[<PC3>=-0.08 \ \  \sigma_{PC3}=0.05 \]

Thus, the color-space occupied by ``normal'' galaxies (all but AGNs) is almost 
two-dimensional. The first two PCs alone contain 93.8 \% and 4.9 \% of the 
data's variance respectively, and together account for 98.7 \% of the total 
variance. 
 This means that, in practice, a two-dimensional parameter space suffices to 
describe the entire range of Hubble types and the different subgroups therein.  
In fact, one single combination (PC1) is already extremely  comprehensive! The 
ensemble of galaxies has the largest scatter (larger than in any original 
color-plane) in the (PC1, PC2) plane, which makes it useful for distinguishing 
among different types of galaxies, as will be described in section 4. PC3 
exhibits the smallest scatter, its standard deviation is of the order of the 
measurement uncertainties, and its full extent amounts only to 0.22 mag. Thus 
PC3 can be used as a characteristic to separate cluster galaxies from other 
objects, as will be described in sections 3.2 and 3.3.

The three PCA components define, therefore, a three-dimensional volume, which 
contains all the galaxy templates. We call this volume ``the cluster-box'', 
because the only condition we imposed on the SED templates was that they be at 
{\em z}=0. Assuming that the spectral variety of local galaxies is 
representative of the evolution of galaxies from {\em z}=1 to the present, the 
cluster-box ought to be invariant in rest-frame observations of distant cluster 
galaxies.

\subsubsection{Robustness: error estimates}
How are the PCs affected by  measurement uncertainties in the input spectra?
The noise in the four original bands being uncorrelated by definition, implies 
that:
\[ M = S + N \]
where M and S are the correlation matrices of the measured colors and the pure
signal respectively, and N (the correlation matrix of the noise) is diagonal, 
the i-th term on the diagonal being the variance of the noise in the i-th color.
If the variance of the noise is the same for all colors, the eigenvectors of S
and M are identical, i.e., the Principal Components are unaffected by noise. 
However, it is more realistic to assume that the noise in $(u\!-\!v)$ is, say, 
twice as large, as in the two other bands. We have simulated this situation by 
creating 1000  mock sets of template spectra by adding random noise drawn from a 
normal distribution with FWHM=0.1 mag for $u\!-\!v$ and FWHM=0.05 mag for 
$(v\!-\!b)$ and $(b\!-\!y)$, the typical uncertainties in the colors derived 
from the Kennicutt spectra, and performed PCA on each one of these. The median 
deviation angles are 1\arcdeg.3 for PC1, 4\arcdeg.6 for PC2 and  5\arcdeg.4 for 
PC3. The corresponding values and standard deviations for the coefficients of 
the Principal Components are :
\[PC1=0.81(\pm 0.01) (u-v) + 0.52(\pm 0.01) (v-b) + 0.28(\pm 0.01) (b-y) \]
\[PC2= -0.56(\pm 0.02) (u-v) + 0.56(\pm 0.03) (v-b) + 0.61(\pm 0.04) (b-y) \]
\[PC3= 0.16(\pm 0.03)  (u-v) - 0.65 (v-b)(\pm 0.03) + 0.74(\pm 0.03) (b-y) \]
The small scatter proves that the cluster-box is very robust against  
uncertainties in the input spectra. For PC1 and PC2 this is not surprising, 
because it is an intrinsic property of PCA to cancel out uncorrelated variations 
(noise) in the presence of (correlated) variations in signal. Folkes, Lahav \& 
Maddox (1996) have shown that the high order PCs, which are dominated by noise,  
can vary greatly in such simulations. In our case, however, the only 
noise-affected component is PC3, and since we are working in three-dimensional 
space, once PC1 and PC2 are established, the third component is fully 
determined. Thus even PC3 is robust.

\subsubsection{Completeness}
 Next, we want to verify whether the PC space defined above is not too 
restrictive to include all possible galaxy types (our galaxy sample is quite 
conservative, because Kennicutt's sample is restricted to local field galaxies, 
and spatially integrated spectra are otherwise rare in the literature). We 
therefore performed the same analysis on a list of  Str\"{o}mgren colors of 143 
galaxies compiled by one of us (KR). This list includes 63 of the Kennicutt 
galaxies as representatives of normal galaxies, 41 Seyfert galaxies from de 
Bruyn \& Sargent (1978), 17 dust-rich galaxies (IRAS sources)  from Ashby, Houck 
\& Hacking (1992), as well as 37 nearby and distant cluster galaxies from Yee \& 
Oke (1978) and Gunn \& Oke (1975) to avoid any cluster/field or redshift bias 
(see Table 3 for details). Figure 4 summarizes the morphological breakup of the 
three tables in form of histograms. 
This sample covers a wider range of spectromorphological\footnote{The term 
``spectromorphological'' denotes the fact that, although the original Hubble 
typing is based on morphology, its terminology is often applied to the spectral 
sequence of galaxies, because of the reasonable correlation between them.} types 
than the Kennicutt + Kinney et al. list, namely it contains strong AGNs, more 
starbursts, and cluster galaxies  (cD's,...).   Such objects are known to be 
present in distant clusters, therefore we must ensure that the selection 
criterion described above, based on the cluster-box, is not biased against them. 
The disadvantage of the Rakos sample is that it does not always meet the large 
aperture condition. This is namely the case of the IRAS galaxies, the Seyfert 
spectra of de Bruyn \& Sargent, and some of the Yee \& Oke objects. 

The space generated by all normal galaxies, as well as Seyfert 2s, in this 
test-sample is nearly identical to the original PCA-cluster-box. In fact, all 
the galaxies of the Rakos sample, except pure Seyferts and  some IR-bright 
galaxies, fall within the cluster-box defined above. 
In order to retain the aperture criterion, we will define the galaxy-PCA-space 
as the union of two subspaces,
one of normal galaxies - as  defined at the beginning of this section - and one 
of galaxies with active nuclei (Seyfert-box), as displayed in Figure 5. In the 
case of the latter, aperture effects are less important because most of the 
emitted light originates from a small nuclear region.
Plotted also in Figure 5 are the IR-bright galaxies from Ashby et al. (1992). 
Some of them lie outside the cluster-box and are therefore visible (as x-es) in 
Figure 5. Visual inspection of their spectra reveals that they do not represent 
a distinct category of galaxies, but rather the highly dust-reddened versions of 
mixed AGN-starburst galaxies.

In summary, an object is identified as cluster member if it lies within either 
of the two boxes defined above. In a second step, one can "redden" the two boxes 
to search for dust-enshrouded cluster members. This makes our selection 
criterion essentially free of any evolutionary
bias. There remains a possibility  that a distant cluster might harbor
a galaxy so peculiar that it resembles nothing we know from the local
samples. We accept, as a caveat, that the membership selection might
reject very few and very peculiar objects, but we consider such a
possibility very unlikely and estimate that it will not influence
significantly our evolutionary conclusions.

\subsection{Star contamination}

Images of galaxy clusters will always be contaminated by foreground stars.
Bright stars are easily identified and discarded, but faint ones can be mistaken 
for  small, compact galaxies at high redshift, especially in observations with 
degraded spatial resolution.

The rest-frame observing technique can considerably alleviate this problem.
The rest-frame ({\em z}\/=0.2 to {\em z}\/=0.8) Str\"{o}mgren filters will 
sample the {\em red} portion of the stellar ({\em z}\/=0) spectra, away from the 
distinctive features they were originally designed to match. This yields 
unrealistic  stellar colors.

In order to investigate this further, we simulated stellar contamination by 
folding the theoretical flux distributions from the 
lcb97\footnote{http://www.astro.unibas.ch/${\rm\sim}$lejeune} library of stellar 
atmospheres (Lejeune, Cuisinier \& Buser 1997, 1998), covering spectral types O 
- M and luminosity classes I, III and V, with the redshifted Str\"{o}mgren 
filter response curves for $0.1 \leq z_{obs} \leq 0.8$. The lcb97 is the most 
complete stellar library to date, covering  $2000 K \!\leq \! T_{eff}\! \leq \! 
50,000K$, $-1.02\! \leq \! \log \! \leq \! 5.5$, and $-5.0\! \leq \! log 
Z/Z_{\sun}\! \leq \!+1.0$ in a uniform grid. In the three-dimensional PC-space 
no star lies inside the Seyfert-box and most stars lie well outside the 
cluster-box (empty space in Figure 6). Figure 6 shows only those stellar types 
that are indistinguishable from galaxies in the redshift range $0.1 \!\leq \! 
z_{obs}\! \leq \! 0.8$. The conclusion is that in observations of clusters at 
any given redshift, only a certain subgroup of F, G,  and K stars will 
contaminate the data. The precise spectral types of the intruding subgroup will 
depend on the redshift of the target cluster.

The actual number of such stars depends on galactic coordinates. This is dealt 
with statistically, as follows. For each cluster, the Bahcall-Soneira Galaxy 
model\footnote{http://www.sns.ias.edu/${\rm \sim}$jnb/Html/galaxy.html} (Bahcall 
1986) is used to predict the number of such stars along the line of sight. This 
model of the Milky Way is the combination of a disk and spheroid, each with its 
own luminosity function, that predicts the projected stellar number density vs. 
apparent magnitude and  B-V color for given Galactic coordinates, as well as the 
breakdown between dwarf and giant stars. Despite its very simple assumptions, 
the model fits amazingly well the observed star counts (Bahcall \& Soneira 
1984).

This is best illustrated by an example : Cl0016+16, a cluster at {\em z}\/=0.54, 
has been observed by KR at KPNO. At that redshift, only stellar types F8I-K2I, 
F2III-K3III and F2V-K4V in the magnitude range m(V)=18.7 - 22.5 (brightest 
cluster galaxy and limiting magnitude of the observation, respectively) can 
contaminate the galaxy counts. Figure 7 shows the projection of the sheet 
containing the stellar  models onto the first  two PC's. In this case, the field 
of view was $7 \times 7$ square arcminutes. The Bahcall-Soneira program predicts 
about 15 such stars within the frame. This is an upper limit, as the 
incompleteness at faint magnitudes has not been folded into the calculation.
For our observations of A115 ({\em z}\/=0.191, Rakos et al. 2000), covering a 
circular region of 10 arcminutes in diameter, 2.6 stars brighter than m(V)=21.25 
(faintest object detected) are predicted to contaminate the cluster-box. This 
calculation takes the incompleteness function into account.

Since - as will be shown in section 4 - different galaxy types occupy distinct 
regions of the cluster-box, we also know to which class of galaxies the stellar 
contamination correction needs to be applied, according to the position of these 
stars {\em within} the cluster-box.

Contamination by stars in the region of highly reddened cluster galaxies is not 
an issue, as stars have low PC2 values (see Fig. 7), whereas reddened objects 
have increasingly high PC2 values (see reddening vector in Fig. 10).

\subsection{Redshift discrimination}

Probably {\em the} most severe weakness of cluster photometry is contamination 
by fore- and background galaxies, especially when trying to investigate the 
outskirts of clusters and/or its faintest members. Considerations of apparent 
size and magnitude do not help to resolve this issue, as both quantities change 
rather slowly with redshift beyond {\em z}\/=0.2, while the intrinsic scatter at 
any given redshift is very large.
The only constraint they yield, is that galaxies brighter and/or larger than the 
first-ranked cluster galaxy can be immediately discarded as foreground, but the 
density of such field galaxies is so low that it hardly improves the situation. 
For the purpose of the following simulations, we have adopted $ 
M_{V}(brightest)\!=\!-22.68 $ (Hoessel, Gunn \& Thuan 1980), $H_{o}\!=\!60$, and 
$q_{o}\!=\!0.5$. When dealing with real cluster data, we will replace this by 
the observed value.

Connolly et al. (1995) have demonstrated the efficiency of photometric redshift 
determinations, which rely mainly on the shifting of the 4000\AA\, break through 
the successive filter passbands. Along the same line of thought, Fiala, Rakos \& 
Stockton (1986) have used the marked difference in their $b\!-\!y$ and {\it mz} 
indices to identify cluster ellipticals at various redshifts.
However, the situation becomes less clear when dealing with the whole gamut of 
galaxy types, as intrinsic spectral differences become entangled with 
differences due to redshift.
For early type galaxies, the shifting (with redshift) of the 4000 \AA \, break 
through the filter bandpasses makes cluster membership identification relatively 
easy, but this criterion becomes useless for active galaxies with flat spectra, 
as well as for any foreground galaxy, that is sampled only redward of the break. 
In the latter cases, one has to rely on a combination of more subtle features, 
such as emission and absorption lines, increasing the dimensionality of the 
problem.
Here again,  PC-space offers the most efficient cure.

In order to assess the efficiency of rest-frame Str\"{o}mgren photometry at 
identifying, and subsequently discarding, interlopers from the list of cluster 
galaxies, we have artificially redshifted the Kinney galaxy template spectra 
from $z_{gal}\!=\!0.0$ to $z_{gal}\!=\!0.80$ in steps of $\delta z \!=\!0.005$, 
to simulate field galaxies. We then simulated rest-frame cluster observations 
over that same redshift range ($z_{obs}$) and calculated for each template the 
PC-colors at every point in the ($z_{gal}$, $z_{obs}$) space. If the PC-colors 
of the redshifted template are inconsistent with the cluster-box, the field 
galaxy will be successfully identified as an interloper. The results are 
(pessimistically) summarized in Figure 8, which shows four representative 
examples. For each galaxy type we shaded that area in the ($z_{gal}$, $z_{obs}$) 
plane where the galaxy's PCs lie {\em inside} the cluster-box. The patch along 
the principal diagonal 
($z_{gal}\!=\!z_{obs}$) represents the {\em correct} identification of cluster 
members, the other patches are {\em failures} of the system to recognize field 
galaxies. Success is demonstrated by the emptiness of the figure. Photometric 
uncertainties (typically 0.05 mag) have been included in this calculation and 
make the interloper strips twice as wide as they would be in the ideal (zero 
error) case. The main conclusion is that our system recognizes background 
galaxies extremely well: only a small fraction of background starburst galaxies 
are not recognized. Foreground galaxies are more problematic, many will be 
mistaken as cluster members. This is not dramatic for low and intermediate 
redshift observations, because the foreground volume subtended by the image size 
is  rather small, but it can be a problem for high-z observations. Luckily, this 
will be somewhat balanced by the smaller magnitude range (to the detection 
limit) spanned by cluster members at high z. It is worth noting that 70 \% of 
the successful identifications are due to PC3 alone. Indeed, as stated in the 
previous section, the allowed range in PC3 for normal galaxies is very small and 
there is no visible trend of rest-frame-PC3 vs. galaxy type. This means that PC3 
is a characteristic quantity for {\em any} kind of galaxy (except AGNs) at the 
right redshift.

Since our field galaxy identifier is not perfect, we estimate in what follows 
the actual number of interlopers remaining after the rejection procedure.
Approximately 15 \% of local field galaxies are currently undergoing starbursts 
(T. Contini, private communication). The remaining 85 \% can be split according 
to their morphology into 8/10 spirals, 1/10 S0s and 1/10 ellipticals (Sandage \& 
Tamman 1979). This yields a spectrophotometric breakup of 15 \% starburst 
galaxies, 68 \% spirals and 17 \% ellipticals and S0s at z=0. We estimate the 
redshift evolution of the spectromorphological breakup based on the luminosity 
density evolution estimated by Lilly et al. (1996). They find that the 
contribution from galaxy types Sbc or later roughly doubles by z=0.75-1.00 in 
the 4400 \AA \, as well as in the one~micron bands. Assuming a one-to-one 
relationship between luminosity density and number counts, we estimate that the 
fractional contribution of our starburst category also doubles, increasing to 
about 30 \% at $z\!=\!1$ at the expense of the other types.
Thus, we can calculate the fraction of field galaxies that our algorithm will 
recognize as a function of $z_{obs}$ and $z_{gal}$. Note that the zero-redshift 
estimate of the spectromorphological breakup takes
into account the typical completeness limits of our observations ($I_{AB}=20-21 
$ mag), whereas its redshift-evolution is derived from the CFRS, which is 
complete to $I_{AB}=22.5$. This  causes us to somewhat overestimate the number 
of high redshift, star-forming galaxies visible in our frames, due to
the luminosity-dependence of galaxy evolution. However, as long 
as $z_{obs} < 0.7$, the field galaxy rejection mechanism works at nearly 100\% 
efficiency for high redshift ($z_{gal} > 0.7$) galaxies (fig. 8 and 9), 
eliminating this 
uncertainty from the end result "number of unidentifiable field galaxies".
For $z_{obs} > 0.7$, the numbers quoted are upper limits.

 Figure 9 shows the percentage of galaxies whose colors are within the 
cluster-box boundaries. The peak along the principal diagonal is produced by the 
cluster galaxies themselves. Success at identifying interlopers corresponds to 
low values in this figure. In order to explain the width of the peak, Figures 
10a-10c show - for $z_{obs}=0.1$, $0.4$ and $0.7$ respectively - the effect of 
small redshift deviations ($-0.1 \leq \delta z \leq 0.1$) about the cluster 
mean. The details depend on galaxy type and redshift of observation. Overall, 
background objects
are more easily discarded than foreground, and early type galaxies more easily 
than late types. The regions of redshift space, where immediate fore- or 
background galaxies still fall within the cluster-box boundaries or into the 
Seyfert-box (see stb3 in Fig. 10a) are accounted for statistically.

We choose to normalize the fraction displayed in Figure 9 by the statistically 
complete spectroscopic subset of the Canada-France Redshift Survey (Lilly et al. 
1995; Crampton et al. 1995) to obtain actual numbers. Its completeness limits 
($17.5 \leq I_{AB} \leq 22.5$) are consistent with our deepest cluster 
observations. Each cluster observation is limited by the flux of objects in its 
respective {\em yz} filter, which corresponds to {\em rest-frame} $m_{5500}$. 
Therefore each {\em yz} filter ``sees'' a different portion of the field 
galaxies' SED (depending on $z_{gal}$), which implies that the number of field 
galaxies seen in cluster observations varies with $z_{obs}$. The same procedure 
can be applied to the Seyfert-box separately. Figure 11 displays the number 
density of remaining interlopers in the cluster-box (pluses) and the Seyfert-box 
(x-es) in three magnitude bins, as a function of $z_{obs}$. The efficiency of 
our algorithm in rejecting interlopers varies between 50 and 100 \%, depending 
on the magnitude bin and redshift of observation. Note that this statistical 
estimate does not take into account the unknown, large-scale clustering features 
of the observed fields.

Contamination of the reddened boxes can be calculated the same way.
Reddened Seyfert-boxes are essentially free of contamination, as well as 
cluster-boxes for $A_{V} > 1.5$. Less reddened cluster-boxes suffer very small 
contamination.

Our goal is to detect evolution in cluster galaxies. Hence the
question of interest is: What is the weakest evolutionary trend we can
positively detect, in spite of the remaining interlopers ? The answer
depends not only on $z_{obs}$, but also on the overdensity in a given
cluster region. In the central regions of rich clusters, field
contamination is not an issue because the overdensity can reach a few
hundred, but previous studies (Rakos, Odell \& Schombert 1997)
indicate that the outer regions are of interest, because the
preferred location of the star-forming galaxies constitutes an
important clue to the physical processes at work. Figure 12 shows the
fraction of unidentifiable interlopers likely to remain in our data
sets after the cluster-box test, as a function of $z_{obs}$ and
overdensity. As long as the overdensity is larger than 10, our worst
case of field galaxy identification (50 \%) is sufficient to remove
any uncertainty introduced by the field galaxies. And even in regions
only twice as dense as the field, the remaining contamination is as
little as 10 \%  at $z\!=\!0.1$ and 30 \% at $z\!=\!0.8$! Since we can
also determine the location within the cluster-box of these
interlopers, we can make very precise corrections to our cluster
statistics.

\subsubsection{Cluster velocity dispersion}
The peculiar velocities of galaxies in clusters can be very large,
velocity dispersions of 1000 - 1500 ${\rm km \, s^{-1}}$ are commonly cited in 
the
literature. Lately, smaller values tend to be published, based on
the virialized population only and where substructures have been
separated, but these considerations do not apply to our study, as we
are interested in infalling galaxies as well and prefer to regard a
recent merger of two sub-clusters as one unit.

Our filters are wide enough to accommodate the red- and blue-shifts
introduced by these large velocities without any substantial change in
colors, i.e., a genuine cluster member with a peculiar velocity of
1500 ${\rm km \, s^{-1}}$ or even 3000 ${\rm km \, s^{-1}}$ will not be 
mistakenly discarded as an
interloper. The most significant difference in PC-colors between a galaxy at
rest and one with a velocity of $\pm$ 1500 ${\rm km\, s^{-1}}$ is 0.0146 in PC1 
for an S0 ( 44\% of the
measurement uncertainty), but typical values are much smaller, of the
order of 0.004 - 0.012 (10\%-15\% of the measurement uncertainties). For a
velocity of $\pm$ 3000 ${\rm km \, s^{-1}}$ these values roughly double, but 
such
objects are much rarer. We conclude therefore, that the width of our
filters is well adapted to the cluster velocity dispersion and that
peculiar velocities do not bias our classification.

\section{Physical interpretation of the Principal Components}
 So far, we have focused on defining the ensemble of cluster galaxies as a whole 
vs. other objects that are likely to populate the parameter space. The next step 
is to ``zoom in'' on this ensemble, in order to characterize differences among 
cluster members themselves. The preceding discussion is entirely based on purely 
observational quantities and empirical relations. We will now show how the same 
quantities can be used to differentiate between different types of cluster 
galaxies. The aim of this section is to outline briefly the link between the 
Principal Components and the underlying physical properties of galaxies. In a 
follow-up publication, we will examine in greater detail the behavior of 
PC-colors as a function of physical properties of galaxies (age, metallicity, 
internal extinction, mixed stellar populations,...) via population synthesis 
models. 

\subsection{PC1: The Hubble Sequence}
The first Principal Component, PC1, follows essentially the Hubble sequence, 
from Ellipticals with the highest positive values of PC1 to Irregulars with the 
most negative PC1s. We have tested this assumption on the ``normal'' galaxies in 
our sample, i.e., the Kennicutt galaxies for which no nuclear activity or any 
signs of interaction with another galaxy has been detected (Figure 13). We have 
relied on the NED database for morphological classification. The statistics of 
this sample being rather poor, as only one third of the Kennicutt galaxies meet 
the above criteria, we use population synthesis models to firm up our statement. 

Old stellar populations are represented by G. Worthey's 
models\footnote{http://199.120.161.183/${\rm\sim}$worthey/dial/dial\_a\_model.ht
ml} (1994). Figure 14 shows that the location of single burst populations viewed 
after 1 to 18 Gyrs (from left to right) are consistent with the location of 
early type galaxies. Only a solar metallicity burst (pluses) and a 
$Z/Z_{\sun}\!=\!-2$ model (x-es) are shown in the figure for clarity. Other 
sub-solar metallicity bursts, or superpositions of bursts, yield similar 
results.
We use the Starburst99\footnote{http://www.stsci.edu/science/starburst99} models 
(Leitherer et al. 1999) to model young stellar populations. The region occupied 
by late type galaxies (left side of the cluster-box) can be fitted with 
moderately reddened bursts of star formation (Fig. 15a), as well as with 
continuous star formation (Fig. 15b), viewed at least 4 Myrs after the initial 
burst. The figures show only the tracks for solar metallicity and a Salpeter 
IMF, but the situation is not much different for other metallicities and Initial 
Mass Functions. The tracks range from 1 to 900 Myrs (the age-range provided by 
the models), with approximate ages marked on the figures. The models lie left of 
the cluster-box for the first few million years. This is also the location of 
individual HII regions. We have nevertheless {\em not} extended the cluster-box 
to cover this region, because the unlikely chance to catch a galaxy in such a 
stage with large aperture photometry would be outweighed by the contamination 
from stars and field galaxies that this would falsely include in the cluster. 
Note that the models do not account for an older underlying population, which is 
certainly present in large aperture photometry. It is therefore not surprising 
that many template galaxies lie on neither of the simplistic tracks shown here.
 The Starburst99 models also do not account for metal enrichment or internal 
extinction, and are, therefore, not directly comparable to large aperture galaxy 
photometry. Both effects tend to make the evolutionary tracks more horizontal, 
under the assumption that heavy elements need to be produced (and ejected) 
before they start to affect the SED.

Both observations and models indicate that PC1 increases monotonically with mean 
stellar age. The "loop" at $t \sim 10 $Myrs in Fig. 15a is an artefact of the 
modeling technique. Thus, PC1 is a good indicator of global star formation 
history. By comparing Figure 13 to Figure 10, one can see that large peculiar 
velocities  would not alter the classification, as differential redshift effects 
are mainly perpendicular to PC1, and are generally small. Also shown in Figure 
10 are the internal reddening vectors for $A_{V}=1$. Note that the templates, 
from which the cluster-box was constructed are spatially integrated spectra of 
real galaxies and therefore they are already reddened by the amount of dust 
present in each galaxy. The vectors of "additional reddening" indicate that (a) 
a cluster galaxy that is unusually reddened will come to lie outside the 
cluster-box, but such objects can be specifically searched for (see section 
3.1.2). Late type and Seyfert galaxies, shifted by $A_{V}\sim 1.5-3.0$, would 
resemble the IR-luminous galaxies detected by IRAS. (b) Galaxies with only 
slightly more reddening than is usual for their type are moved along PC1 towards 
earlier types.

\subsection{[PC2,PC3]: Active Galactic Nuclei}
It has already been hinted in section 3.1 (Fig. 5) that most galaxies with 
active nuclei do not fit into the cluster-box. Their total (nucleus + host) PC1 
values are not much different from those of normal galaxies (slightly bluer, but 
we cannot determine on the basis of our sample how much of this is a real trend 
or due to aperture effects), but their location in the [PC2, PC3] plane is 
inconsistent with any kind of normal galaxy. Since the Hubble sequence is 
essentially parallel to PC1, any trend seen in the plane perpendicular to it 
will be independent of Hubble type.

Figure 16 displays all the Seyfert galaxies from the Kennicutt and Rakos samples 
in the [PC2, PC3] plane. The projections of the cluster-box and Seyfert-box 
(defined in section 3.1.2, see Figure 5 for a 3D view)  are also shown. There is 
a clear trend with Seyfert type: Sy2s are indistinguishable from normal 
galaxies, but Seyfert types 1.5 to 1 tend to lie further away from the 
cluster-box, mainly due to their low PC3 values, which are  signatures of the 
nuclear flux. This trend can be interpreted in terms of inclination of the host 
galaxy (Antonucci 1993) determining the relative contribution of the nucleus and 
host galaxy stellar population to the total SED, and it is to be expected that 
aperture effects cause a similar trend.
Thus, PC3 will not detect {\em every} active nucleus (unless the spatial 
resolution is good enough to permit adequate surface photometry) but it will 
surely identify galaxies in which the signature of nuclear activity dominates 
the spectrum (i.e., the object is in the Seyfert-box). Note that contamination 
of the Seyfert-box is very small: For most $z_{obs}$, not even one star is 
likely to reside within it, and the field galaxies (see Figure 11) are also very 
few (at most one, for our observations of three $z \sim 0.2$ clusters).

\subsection{PC3: The age-metallicity degeneracy}
One of the major drawbacks of standard optical broad-band photometry
is the  age-dust-metallicity degeneracy (Worthey 1994; Charlot,
Worthey \& Bressan 1996), that can only be resolved by adding specific
line-indices, at high observing time cost (for the metallicity) and
UV+IR bands for the dust. We found that PCA on rest-frame
Str\"{o}mgren photometry can resolve this degeneracy, if redshift
information is independently available.

Figure 17 displays the models for old stellar populations in the [PC1, PC3]  
plane. In this plane, the dust reddening vector is at a large enough
angle from the line formed by the various ages and metallicities, that
substantial reddening can be distinguished from age and metallicity
effects. The extinction by dust was calculated assuming the reddening law of 
Savage \& Mathis (1979) and a simple intervening layer geometry. We have not 
used other reddening laws, such as LMC- or SMC-type laws, because they do not 
differ from the Milky Way law in the optical domain (Calzetti 1998).  Inspection 
of the reddening laws produced by more complicated geometries, such as unevenly 
distributed, clumpy dust (Calzetti 1997) suggests that the main difference will 
be only in the normalization, and not in the orientation of the reddening 
vector. We can thus lift the dust-degeneracy in galaxy photometry, provided the 
nature and distribution of the dust, derived from nearby objects, holds at least 
for $z\!\leq\!1$.

In Figure 18 we have plotted the evolutionary tracks of old stellar
populations for different metallicities  in the [PC2, PC3] plane. None
of them overlap, which means that in principle, the age-metallicity
degeneracy is resolved in this set of coordinates. In practice, our
typical measurement uncertainties ($\sim 0.05$) will allow us to make
only qualitative metallicity statements, except for the brightest
cluster galaxies, for which 1 \% photometric accuracy is
achievable. Again, this requires prior knowledge of the galaxy's redshift at the 
 percent level (see Fig. 10). \\

The combination of these two figures enables us to disentangle the effects of 
age, metallicity, and dust for old ($\geq 1 $Gyr), single stellar populations, 
based on only three optical colors.

\section{Summary}
We have shown how rest-frame photometry using intermediate-band filters, an 
observing technique developed by KR, fills the gap between standard photometric 
and spectroscopic observations of galaxy clusters. The advantages of this method 
are:
\begin{itemize} 
\item Large fields of view can be acquired in a single telescope pointing and 
all sources present can be observed and analyzed simultaneously (the usual 
advantages of photometry). 
\item The use of filters {\em tuned} to the redshift of each cluster largely 
overcomes the usual problem of k-corrections, allowing direct comparison of data 
over arbitrary large redshift ranges, without any a priori theoretical 
assumptions.
\end{itemize}
In this paper, we have systematized the analysis of such data using the 
technique of Principal Component Analysis. We have applied PCA to synthetic 
Str\"{o}mgren colors of a sample of well-known, nearby galaxies covering as much 
as possible all galactic properties. We demonstrated that PCA:
\begin{itemize}
\item Finds the orthonormal coordinate space in which the data are most simply 
described. This makes it very easy to fully automate all subsequent analysis.
\item Improves and separates the information content of the data. Combined with 
the wise choice of the filter's rest-frame central wavelengths, we thus gain a 
considerable amount of spectral information without recourse to actual  
spectroscopy.
\end{itemize}

Based on a training set of nearby galaxies, we have shown that we are able to:
\begin{itemize}
\item Substantially reduce contamination by field stars and fore- or background 
galaxies,
\item Characterize the global star-formation histories of cluster galaxies, and
\item Identify AGN activity of cluster members (via the Seyfert-box).
\end{itemize}

\begin{itemize}
\item By analyzing models of stellar populations in the same manner, we have 
resolved the age-dust-metallicity degeneracy for old ($\geq 1$ Gyr.) stellar 
populations. This promises new insights on the cosmological evolution of dust 
and metallicity, when applied to cluster ellipticals at various redshifts.
\end{itemize}
We believe that the ratio of information to observing and reduction time 
required for the application of this method makes it an ideal tool for cluster 
galaxy surveys, because it allows one to address simultaneously many 
evolutionary issues, such as dependence on redshift as well as on environment 
(cluster type, location within a cluster, ...).

\acknowledgements
Many thanks to Ofer Lahav for sharing his experience in PCA and to Thierry 
Contini for discussions on starburst galaxy statistics. We are grateful to H. 
Netzer for helping us classify AGN and IRAS spectra and reading so many drafts 
of this paper. We also express our gratitude to the referee, David Koo, for 
constructive remarks that improved this paper. 

This research has made use of the NASA/IPAC Extragalactic Database (NED) which 
is operated by the Jet Propulsion Laboratory, California Institute of 
Technology, under contract with the National Aeronautics and Space 
Administration. Research at the Wise Observatory is supported in part by a grant 
from the Israel Science Foundation. S. Steindling was supported partly by a 
grant from the US-Israel Binational Science Foundation.

\newpage
\section*{Figure captions}
\figcaption[]{{\bf Three-dimensional view of the Kennicutt and Kinney galaxies 
in classical Str\"{o}mgren color-space.} The galaxies occupy a well-constrained 
subspace, inside which they are ordered roughly according to Hubble type, with 
early types in the upper right corner. The projections onto the three sub-planes 
are also shown in gray. As explained, such color coordinates are impractical for 
a quantitative description.  circles: E and S0, triangles up: Sa - Sb, triangles 
down: Sc - Sd, diamonds: Im, squares: I0, pluses: unspecified Spiral, stars: the 
Kinney et al. starburst templates. The classification is taken from NED. Open 
symbols: galaxies with reported Seyfert(2) nucleus, filled symbols: no
reported Seyfert activity.}

\figcaption[]{{\bf Galaxies in PCA-space.} The Kennicutt+Kinney sample viewed in 
3D PCA-space. The three projections onto the PC1-PC2, PC2-PC3 and PC1-PC3 planes 
are also shown in gray. Differences among galaxies now spread almost solely 
along one axis, PC1. High values of PC1 correspond to early type galaxies, low 
values to late types. The dispersion along PC2, and even more so along PC3, is 
much smaller. The standard deviation in PC3 is of the order of the measurement 
uncertainties in the data. This means, that ${\rm <PC3>}$ is an intrinsic galaxy 
``property'', which can be used to distinguish the ensemble of cluster galaxies 
from other objects, such as stars and field galaxies. The advantage of using 
PC-space instead of color-space is evident. The cluster-box, as defined in the 
text, and its projections are also shown. Symbols as in Figure 1.}

\figcaption[]{{\bf Second order k-corrections of PC colors.} Filters with 
central wavelengths tuned to fixed rest-frame spectral regions of galaxies at 
different redshifts, but of constant width, sample narrower bands of the 
galaxie's spectrum with increasing z. Panels (a), (b), and (c) show the 
differences in PC1, PC2 and PC3 due to this effect. Different symbols are used 
for each template spectrum. The dash-dotted lines indicate the uncertainties 
corresponding to d(u-v)=d(v-b)=d(b-y)=0.05.}

\figcaption{{\bf Morphological breakup of the three sets of reference spectra.} 
The upper panel is the Kennicutt sample, the middle panel  shows the individual 
Kinney et al. galaxies, and the lower panel displays the Rakos sample. Galaxies 
too peculiar to be assigned to any Hubble type are assigned T=15 and galaxies 
with no morphological entry neither in the source paper nor in NED are plotted 
with T=17.}

\figcaption[]{{\bf Testing for Completeness.} The large box represents the 
cluster-box defined in Figure 2. All normal galaxies from the Rakos test sample 
(see text) are located within this box and are therefore not visible in the 
plot. The smaller box below it encompasses the location in PC-space of Seyfert 
nucleus dominated galaxies. These two volumes do not overlap. The location of 
weaker Sys, which have a composite spectrum of AGN + host galaxy, can be 
explained by a simple combination of ``pure'' Seyfert and ``pure normal galaxy'' 
(see section 4 for a more detailed discussion).
Also shown is the internal reddening vector for $A_{V}\!=\!1$ (solid arrow) and 
its projections along each axis (dashed arrows), and the IR-bright galaxies 
(x-es). Such objects do not represent a separate class of galaxies but reddened 
versions of mixed AGN-starburst galaxies (again, see section 4).}

\figcaption[]{{\bf Contamination by stars} for observations with $0.1 \leq 
z_{obs} \leq 0.8$. The figure displays those stellar types that will be 
indistinguishable from cluster galaxies, based on their colors alone. The 
efficiency with which we can eliminate foreground stars is reflected by the 
emptiness of this figure.}

\figcaption[]{{\bf  Stellar contamination in Cl0016+16.} Projection onto the 
(PC1, PC2) plane of the cluster-box and the stellar sequence. Different symbols 
are used to distinguish spectral types, and luminosity class is represented by 
the symbol size (large symbols for supergiants, intermediate size for giants,  
and small symbols for main sequence stars). Circles = O stars; pluses = B; x-es 
= A; triangles up = F; squares = G; triangles down = K, and diamonds~=~M.}

\figcaption[]{{\bf Redshift discriminating ability.} For selected galaxy  
templates, we show, for each redshift of observation ($z_{obs}$), the redshifts 
$z_{gal}$ at which this type of field galaxy will be indistinguishable from 
cluster members. The central strips, along the principal diagonal, correspond to 
galaxies for which $z_{gal}\!=\!z_{obs}$, i.e., cluster members. All quiescent 
background galaxies (upper panels) are successfully discarded, as are most 
starbursts.}

\figcaption[]{{\bf Redshift discriminating ability.} Percentage of galaxies, as 
a function of $z_{obs}$ and $z_{gal}$, whose PC-colors are within the boundaries 
of the cluster-box. The peak along the principal diagonal corresponds to the 
cluster itself.}

\figcaption[]{{\bf Differential redshift effects.} Figs. 10a-10c show - for 
$z_{obs}=0.1$, $0.4$ and $0.7$ respectively - the cluster-box and Seyfert-box 
boundaries, as well as the location of three representative template galaxies 
(E, Sb and stb3 of Kinney et al.) in all three projections (PC1-PC2, PC2-PC3 and 
PC1-PC3). For each template, the tracks for $-0.1 < \delta z < +0.1$ and the 
intrinsic reddening vector for $A_{V}=1$ are shown. Dotted lines: blueshift, 
dashed lines: redshift. Note how the tracks change with $z_{obs}$.}

\figcaption[]{{\bf Redshift discriminating ability.} Number density of visible 
field galaxies (solid line) as a function of $z_{obs}$ and remaining interlopers 
in the cluster-box (pluses) and in the Seyfert-box (x-es) for three apparent 
magnitude bins.}

\figcaption[]{{\bf Weakest detectable trend.} Percentage of unidentified 
interlopers in deep cluster frames as a function of overdensity and cluster 
redshift from $z_{obs}=0.1$ (lowest solid line) to $z_{obs}=0.8$ (upper dotted 
line) in steps of $\delta z =0.1$ (solid and dotted lines are used alternatively 
 for clarity). In the least favorable case ({\em z}\/=0.8, overdensity=2) the 
remaining field contamination (after the cluster-box test) is at most 30 \%. 
Assuming a worst-case scenario, where all interlopers look alike in PC-space, 
any cluster galaxy property shared by more than 30 \% of all objects can still 
be detected. At $z\!=\!0.1$, the weakest detectable trend need only be 10 \%, 
and for denser regions (at any redshift), the statistical uncertainty due to 
remaining interlopers will be completely negligible.}

\figcaption{{\bf The Hubble Sequence in PC-space}} The Hubble Sequence is 
essentially parallel to PC1. High values correspond to early Hubble types, low 
values to late types. There is no visible trend of PC2 (and PC3) with Hubble 
type.  The cluster-box is drawn in a solid line, and the dash-dotted line is 
indicative of the typical 0.05 mag photometric accuracy.
\figcaption{{\bf Models of old stellar populations} (Worthey 1994). 
Instantaneous star formation bursts with solar (pluses) and sub-solar (x-es) 
metallicities, viewed after 1-18 Gyrs of passive evolution, are consistent with 
the location of Elliptical and S0 galaxies. Superpositions of bursts (not shown) 
look very similar.}
\figcaption{{\bf Models of young stellar populations} (Leitherer et al. 1999) 
Solar metallicity tracks with a Salpeter IMF are shown for ages from 1 to 900 
Myrs. (a) Instantaneous burst, (b) continuous star formation. $A_{V}\!=\!1$ is 
the intrinsic reddening vector. Solid rectangle: cluster-box, dash-dotted 
rectangle: cluster-box with typical errors, dashed rectangle: Seyfert-box.}
\figcaption{{\bf Perpendicular to the Hubble sequence: Identification and 
classification of Seyfert galaxies.} Although the Seyfert subgroups are not 
defined consistently with the same precision, a trend (of mainly PC3) with 
Seyfert type is apparent, from Sy1s (pluses) having low PC3s (below the 
cluster-box) to Sy2's (x-es) which are indistinguishable from normal galaxies. 
Circles are Sy1.2 and star symbols are Sy1.5. Typical errors are 0.05 in PC2 and 
0.06 in PC3. The ``normal'' cluster-box is outlined by a solid line, the 
Seyfert-box by a dash-dotted line.}

\figcaption{{\bf Measuring dust extinction.} In this projection, the 
evolutionary tracks are aligned and the dust reddening vector makes a 
sufficiently large angle with them to detect departures from the tracks due to 
reddening. The different symbols represent different metallicities: circles: 
$\log Z/Z_{\sun}\!=\!-\!2$, stars: $\log Z/Z_{\sun}\!=\!-\!1$, x-es: $\log 
Z/Z_{\sun}\!=\!0$. The sub-solar metallicity tracks range from t=8 to t=18 Gyrs, 
the solar track covers t=1 to t=18 Gyrs. These are the age ranges available at 
the model's website.}
\figcaption{{\bf Age vs. metallicity in PC-space.} Evolutionary tracks for 
varying metallicities of old stellar populations. The range of ages of the 
tracks (in Gyrs) is indicated in the figure. Each time-step is 1 Gyr. The 
symbols are as in Fig. 17.}

\begin{deluxetable}{lllllll}
\footnotesize
\tablecaption{The Kennicutt galaxy sample}
\tablehead{\colhead{Name} & \colhead{Kennicutt Morph.} &\colhead{NED Morph.} &  
\colhead{u-v}&\colhead{v-b}&\colhead{b-y}&  \colhead{T-type}}
\startdata
NGC1275  &    Epec   & cD pec Sy2 & 0.31 & 0.08 &  0.34     & -4 \nl
NGC3379  &    E0     &  E1        & 0.96 & 0.41 & 0.34      & -5 \nl
NGC4472  &    E1/S0  &  E2/S02    & 0.88 & 0.43 & 0.41      & -3 \nl
NGC4648  &    E3     &  E3        & 0.98 & 0.37 & 0.34  & -5 \nl
NGC4889  &    E4     &  cD/Db     & 1.00 & 0.38 & 0.29   & -4 \nl
Mrk270   &    S0     &  S0? Sy2   & 0.79 & 0.27 & 0.22  & -2 \nl
NGC3245  &    S0     &  SA(r)0 HII LINER & 0.88 & 0.33 & 0.38  & -2 \nl
NGC3516  &    S0     &  (R)SB(s)0 Sy1.5  & 0.38 & 0.11 & 0.20  & -2 \nl
NGC3921  &    S0pec  &  (R')SA(s)0/a pec & 0.50 & 0.00 & 0.22  & 0 \nl
NGC3941  &    SB0/a  &  SB(s)0    & 0.91 & 0.33 & 0.23  & -2 \nl
NGC4262  &    SB0    &  SB(s)0-   & 0.92 & 0.33 & 0.30  & -2 \nl
NGC5866  &    S0     &  S0        & 0.78 & 0.31 & 0.37  & -2 \nl
NGC1357  &    Sa     &  SA(s)ab   & 0.65 & 0.22 & 0.27  & 2 \nl
NGC2775  &    Sa     &  SA(r)ab   & 0.82 & 0.26 & 0.33  & 2 \nl
NGC3368  &    Sab    &  SAB(rs)ab Sy & 0.75 & 0.29 & 0.32  & 2 \nl
NGC3471  &    Sa     &  Sa           & 0.45 & 0.09 & 0.28  & 1 \nl
NGC3623  &    Sa     &  SAB(rs)a LINER      & 0.84 & 0.37 & 0.30  & 1 \nl
NGC5548  &    Sa     &  (R')SA(s)0/a Sy1.5  & 0.15 & 0.00 & 0.11  & 0 \nl
NGC7469  &    Sa     &  (R')SAB(rs)a Sy1.2  & -0.15 & -0.06 & 0.13  & 1 \nl
NGC1832  &    SBb    &  SB(r)bc             & 0.32 & 0.01 & 0.16  & 4 \nl
NGC3147  &    Sb     &  SA(rs)bc Sy2        & 0.58 & 0.11 & 0.23  & 4 \nl
NGC3227  &    Sb     &  SAB(s) pec Sy1.5    & 0.55 & 0.17 & 0.22  & 1-7 \nl
NGC3310  &    Sbcpec &  SAB(r)bc pec        & -0.08 & -0.44 & -0.03  & 4 \nl
NGC3627  &    Sb     &  SAB(s)b Sy          & 0.39 & 0.07 & 0.19  & 3 \nl
NGC4750  &    Sbpec  &  (R)SA(rs)ab LINER   & 0.46 & 0.08 & 0.23  & 2 \nl
NGC5248  &    Sbc    &  SAB(rs)bc,HII       & 0.32 & 0.06 & 0.20  & 4 \nl
NGC6217  &    SBbc   &  (R)SB(rs)bc Sy2     & 0.34 & -0.09 & 0.09  & 4 \nl
NGC6764  &    SBb    &  SB(s)bc LINER Sy2   & 0.21 & -0.01 & 0.31  & 4 \nl
NGC2276  &    Sc     &  SAB(rs)c            & 0.15 & -0.15 & 0.05  & 5 \nl
NGC2903  &    Sc     &  SAB(rs)bc HII       & 0.33 & 0.05 & 0.24  & 4  \nl             
NGC3690  &    Scpec  &  SBm? pec (strong int. pair) & -0.09 & -0.15 & 0.23  & 9 
\nl
NGC4631  &    Sc     &  SB(s)d       & 0.05 & -0.21 & -0.05  & 7 \nl
NGC4775  &    Sc     &  SA(s)d       & 0.06 & -0.23 & -0.01  & 7 \nl
NGC6181  &    Sc     &  SAB(rs)c HII & 0.28 & -0.06 & 0.20  & 5 \nl
NGC6643  &    Sc     &  SA(rs)c      & 0.39 & -0.03 & 0.11  & 5 \nl
Mrk478   &    Im     &  compact      & -0.40 & -0.22 & -0.15  & 11 \nl
NGC1569  &    Sm/Im  &  IBm Sy1      & 0.19 & 0.07 & 0.31  & 10 \nl
NGC4194  &    Smpec  &  IBm pec      & 0.19 & -0.14 & 0.21  & 10 \nl
NGC4449  &    Sm/Im  &  IBm          & -0.07 & -0.27 & -0.05  & 10 \nl
NGC4485  &    Sm/Im  &  IB(s)m pec   & 0.06 & -0.27 & -0.01  & 10 \nl
NGC5996  &    SBd    &  SB?          & 0.10 & -0.19 & 0.14  & 1-7 \nl
NGC3034  &    I0     &  I0 HII       & 0.55 & 0.13 & 0.42  & 0 \nl
NGC3077  &    I0     &  I0 pec       & 0.40 & 0.04 & 0.18  & 0 \nl
NGC5195  &    I0pec  &  SB0\_1 pec LINER    & 0.80 & 0.32 & 0.36  & -2 \nl
NGC6240  &    I0pec  &  I0: pec LINER Sy2   & 0.52 & 0.07 & 0.31  & 0 \nl
Mrk35    &    pec    &  BCD/Irr             & -0.28 & -0.25 & 0.05  & 11 \nl
NGC7714  &    Spec   &  SB(s)b: pec HII     & 0.05 & -0.24 & 0.10  & 3 \nl
UGC6697  &    Spec   &  Im:                 & -0.03 & -0.21 & 0.06  & 10 \nl
\enddata
\tablecomments{Column 1 lists the galaxy name, column 2 the morphological type 
as listed in the source paper (Kennicutt 1992a). For completeness and 
consistency, we list in column 3 the morphological classification from NED and 
in column 7 the corresponding T-types. Columns 4-6 list the Str\"{o}mgren colors 
used in the PCA-analysis. A histogram of the morphological breakup is shown in 
Figure 11. Galaxies for which no precise morphological typing could be found are 
counted with equal weight in each morphological bin that their classification 
allows.}
\end{deluxetable}

\begin{deluxetable}{llllllll}
\footnotesize
\tablecaption{Kinney et al. galaxy templates}
\tablehead{\colhead{Template} & \colhead{Name} & \colhead{Morph.} & \colhead{NED 
Morph.} &\colhead{u-v}&\colhead{v-b}&\colhead{b-y}& \colhead{T-type}}
\startdata
E  & NGC1399   &   E1pec    &  E1 pec            & 0.92 & 0.47 & 0.36 &   -5 \nl
   & NGC1404   &   E2       &  E1                 & & & &  -5 \nl
   & NGC6868   &   E2       &  E2                 & & & &  -5 \nl
   & NGC7196   &   E3       &  E:                 & & & &  -5 \nl
S0 & NGC1023   &   SB0      &  SB(rs)0-          & 0.93 & 0.43 & 0.38 &   -2 \nl
   & NGC1553   &   S0pec    &  SA(rl)0 LINER      & & & &  -2 \nl
   & NGC6340   &   SA(s)0/a &  SA(s)0/a LINER     & & & &  0 \nl
Sa & NGC1433   &   SBab     & (R'\_1)SB(rs)ab Sy2  & 0.95 & 0.18 &
0.32 & 2 \nl
   & NGC4594   &   Sa       & SA(s)a LINER Sy2    &  & & & 1 \nl
   & NGC4569   &   SABab    & SAB(rs)ab LINER Sy  &  & & & 2 \nl
Sb & NGC210    &   Sb       & SAB(s)b             & 0.73 & 0.23 & 0.31  & 3 \nl
   & NGC7083   &   Sb       & SAB(rs)c LINER      &  & & & 5 \nl
Sc & NGC598    &   Scd      & SA(s)cd             & -0.30 & -0.35 & -0.11  & 6 
\nl
   & NGC2403   &   Sc       & SAB(s)cd            & & & &  6 \nl
\tableline
stb1 ($E(B\!-\!V)\!\leq\!0.1$) & Tol1924-416 & BCG &   pec HII
& 0.24 & -0.29 & -0.07  &   11 \nl
                         & NGC1510     & BCG &   SA0 pec? HII     & & & &  -2 
\nl
                         & NGC1800     & BCG &   IB(s)m           & &
                         & &  10 \nl
                         & NGC1140     & BCG &   IBm pec: Sy2     & &
                         & &     10 \nl
                         & Mrk66       & BCG &   BCG              &  &
                         & &11,(-6) \nl
                         & NGC7250     & stb &   Sdm?             & &
                         & &    8 \nl
                         & NGC5253     & stb &   Im pec HII       & &
                         & &    10 \nl
                         & Haro15      & BCG &   (R)SB0 pec? HII  & &
                         & &    -2 \nl
stb2 ($0.11\!\leq\!E(B\!-\!V)\!\leq\!0.21$) & UGC9560  & BCG & Pec
& -0.04 & -0.34 & -0.06   & 15 \nl
                                    & NGC3125  & BCG & S HII (pair)
                                    & & & &    1-7 \nl
                                    & 1941-543 & stb & XXX
                                    & & & &    17 \nl
                                    & Mrk357   & BCG & Pair?
                                    SBnuc. Sy1 & & & &    1-7 \nl
                                    & UGCA410  & stb & compact
                                    & & & &    11,-6 \nl
                                    & NGC6052  & stb & pair=Sc+Sc
                                    & & & &    5 \nl
stb3 ($0.25\!\leq\!E(B\!-\!V)\!\leq\!0.35$) & Mrk542   & stb & Im:
&  0.34 & -0.15 & 0.06  & 10 \nl
                                    & NGC3049  & stb & SB(rs)ab
                                    & & & &    2 \nl
                                    & NGC5236  & stb & SAB(s)c HII
                                    & & & &    5 \nl
stb4 ($0.39\!\leq\!E(B\!-\!V)\!\leq\!0.50$) & NGC7673  & stb &
(R')SAc? pec HII & 0.33 & -0.18 & 0.02  &  5 \nl
                                    & NGC7793  & stb & SA(s)d
                                    & & & &    7 \nl
                                    & NGC7714  & stb & SB(s)b: pec HII
                                    & & & &    3 \nl
                                    & Mrk499   & BCG & Im:
                                    & & & &    10 \nl
                                    & NGC5996  & stb & SB?
                                    & & & &    1-7 \nl
stb5 ($0.51\!\leq\!E(B\!-\!V)\!\leq\!0.60$) & NGC4385  & stb &
SB(rs)0+: HII    & 0.28 & -0.15 & 0.12  &   -2 \nl
                                    & 1050+04  & BCG & XXX
                                    & & & &    17 \nl
                                    & NGC6090  & stb &
                                    pair=Sbpec+Sbpec HII  & & && 3  \nl
                                    & IC1586   & BCG & compact HII
                                    & & & &    11 \nl
                                    & NGC6217  & stb & (R)SB(rs)bc Sy2
                                    & & & &    4 \nl
                                    & IC214    & stb & XXX
                                    & & & &    17 \nl
stb6 ($0.61\!\leq\!E(B\!-\!V)\!\leq\!0.70$) & NGC5860  & stb &
pair=E+E         & 0.41 & -0.14 & 0.06  &  -5 \nl
                                    & NGC1313  & stb & SB(s)d
                                    & & & &    7 \nl
                                    & NGC1672  & stb & (R'\_1:)SB(r)bc
                                    Sy2 & & & &  4 \nl
                                    & NGC3256  & stb & pec merger HII
                                    & & & &  15 \nl
                                    & NGC7552  & merger & SA(s)c pec
                                    HII LINER & & & & 5 \nl
\enddata
\tablecomments{Column1 lists the template label used by Kinney et al. and column 
2 gives the galaxies from which the optical part of the template spectrum was 
built. Types stb1-6 are templates of starbursting galaxies with different 
amounts of internal extinction (Calzetti, Kinney \& Storchi-Bergmann 1994). 
Column 3 lists the types given by Kinney et al, column 4 the NED morphology, and 
column 8 the corresponding T-type (see histogram in Figure 11). Galaxies for 
which NED does not have a morphological entry are marked as XXX. Columns 5-7 are 
the Str\"{o}mgren colors.}
\end{deluxetable}

\begin{deluxetable}{lllllll}
\footnotesize
\tablecaption{The Rakos galaxy sample}
\tablehead{\colhead{Name} & \colhead{Morph.} & \colhead{NED Morph.} 
&\colhead{u-v}&\colhead{v-b}&\colhead{b-y}& \colhead{T-type}}
\startdata
NGC3379     &   E0              &  E1          & 0.96 & 0.41 & 0.34 &  -5 \nl
NGC4472     &   E1              &  E2/S02      & 0.88 & 0.43 & 0.41  &     -3 
\nl
NGC4648     &   E3              &  E3          & 0.98 & 0.37 & 0.34  &     -5 
\nl
NGC4889     &   E4              &  cD/Db       & 1.00 & 0.38 & 0.29  &     -4 
\nl
standard-E  &   E               &  XXX         & 0.89 & 0.33 &0.35  &     
17,(-5) \nl
NGC1275     &   Epec-burst      &  cD pec Sy2  & 0.31 & 0.08 & 0.34  &     -4  
\nl
NGC3245     &   S0              &  SA(r)0 HII LINER & 0.88 & 0.33 & 0.38  &     
-2 \nl        
NGC5866     &   S0              &  S0       & 0.78 & 0.31 & 0.37  &     -2 \nl
NGC4262     &   SB0             &  SB(s)0-  & 0.92 & 0.33 & 0.30  &     -2 \nl
NGC3941     &   SB0             &  SB(s)0   & 0.91 & 0.33 & 0.23  &     -2 \nl
Mrk270      &   S0-Sy2          &  S0? Sy2  & 0.79 & 0.27 & 0.22  &     -2 \nl
Mrk3        &   S0-Sy2          &  S0: Sy2  & 0.53 & 0.37 & 0.32  &     -2 \nl
NGC3516     &   S0-Sy1          &  (R)SB(s)0 Sy1.5 & 0.38 & 0.11 & 0.20  &     
-2 \nl
NGC3921     &   S0pec-merger    &  (R')SA(s)0/a pec & 0.50 & 0.00 & 0.22  &     
0 \nl
NGC1357     &   Sa              &  SA(s)ab & 0.65 & 0.22 & 0.27  &     2 \nl
NGC2775     &   Sa              &  SA(r)ab & 0.82 & 0.26 & 0.33  &     2 \nl
NGC3471     &   Sa              &  Sa      & 0.45 & 0.09 & 0.28  &     1 \nl
NGC3623     &   Sa              &  SAB(rs)a LINER & 0.84 & 0.37 & 0.30  &     1 
\nl
NGC3368     &   Sab             &  SAB(rs)ab Sy   & 0.75 & 0.29 & 0.32  &     2 
\nl
NGC5548     &   Sa-Sy1          &  (R')SA(s)0/a Sy1.5 & 0.15 & 0.00 & 0.11  &    
 0 \nl
NGC7469     &   Sa-Sy1-merger   &  (R')SAB(rs)a Sy1.2 & -0.15 & -0.06
& 0.13  &     1 \nl
NGC3147     &   Sb              &  SA(rs)bc Sy2     & 0.58 & 0.11 & 0.23  &     
4 \nl
NGC3227     &   Sb              &  SAB(s) pec Sy1.5 & 0.55 & 0.17 & 0.22  &     
1-7 \nl
NGC3627     &   Sb              &  SAB(s)b Sy        & 0.39 & 0.07 & 0.19  &     
3 \nl
NGC4750     &   Sb              &  (R)SA(rs)ab LINER & 0.46 & 0.08 & 0.23  &     
2 \nl
NGC1832     &   SBb             &  SB(r)bc           & 0.32 & 0.01 & 0.16  &     
4 \nl
NGC5248     &   Sbc             &  SAB(rs)bc HII     & 0.32 & 0.06 & 0.20  &     
4 \nl
NGC6764     &   SBb-Sy2-WR      &  SB(s)bc LINER Sy2 & 0.21 & -0.01 & 0.31  &    
 4 \nl
NGC6217     &   SBbc-burst      &  (R)SB(rs)bc Sy2   & 0.34 & -0.09 & 0.09  &    
 4 \nl
NGC3310     &   Sbc-merger      &  SAB(r)bc pec      & -0.08 & -0.44 &
-0.03&     4 \nl
NGC4775     &   Sc              &  Sc           & 0.06 & -.023 & -0.01  &     5 
\nl
NGC6181     &   Sc              &  SAB(rs)c HII & 0.28 & -0.06 & 0.20  &     5 
\nl
NGC6643     &   Sc              &  SA(rs)c   & 0.39 & -0.03 & 0.11  &     5 \nl
NGC2276     &   Sc-merger       &  SAB(rs)c  & 0.15 & -0.15 & 0.05  &     5 \nl
NGC2903     &   Sc-burst        &  SAB(rs)bc HII & 0.33 & 0.05 & 0.24  &     4 
\nl
NGC4631     &   Sc-burst        &  SB(s)d & 0.05 & -0.21 & -0.05  &     7 \nl
NGC3690     &   Sc-merger       &  SBm? pec (strong int. pair) & -0.09
& -0.15 & 0.23&  9 \nl
NGC5996     &   SBd-burst       &  SB?    & 0.10 & -0.19 & 0.14  &  1-7 \nl
NGC4449     &   Sm-merger       &  IBm    & -0.07 & -0.27 & -0.05  &  10 \nl
NGC4194     &   Sm-merger       &  IBm pec & 0.19 & -0.14 & 0.21  &  10 \nl
NGC6052     &   Sm-merger       &  Pair=Sc+Sc & -0.10 & -0.27 & 0.10  &  5 \nl
NGC1569     &   Sm-burst        &  IBm Sy1    & 0.19 & 0.07 & 0.31  &  10 \nl
NGC4485     &   Sm-merger       &  IB(s)m pec & 0.06 & -0.27 & -0.01  &  10 \nl
NGC3034     &   I0              &  I0 HII     & 0.55 & 0.13 & 0.42  &  0 \nl
NGC5195     &   I0pec           &  SB0\_1 pec LINER  & 0.80 & 0.32 & 0.36  &  -2 
\nl
NGC3077     &   I0-burst        &  I0 pec & 0.40 & 0.04 & 0.18  &  0 \nl
NGC6240     &  I0-merger        &  IO: pec LINER Sy2 & 0.52 & 0.07 & 0.31  &  0 
\nl
UGC6697     &   S               &  Im: & -0.03 & -0.21 & 0.06  &  10 \nl
NGC4670     &   SB-BlComGa-burst & SB(s)0/a pec: & 0.10 & -0.23 & -0.07  &  0 
\nl
Mrk35       &   pec-BlComGa-burst & BCD/Irr & -0.28 &-0.25 & 0.05  &   11 \nl
NGC7714     &   Spec-burst      &  SB(s)b: pec HII & 0.05 & -0.24 & 0.10  &  3 
\nl
3C31        &   cD              &  SA0-: & 0.83 & 0.50 & 0.37  &  0 \nl
3C33        &   cD              &  pair  & 0.59 & 0.29 & 0.22  &  17 \nl
3C76        &   cD              &  E1?   & 0.78 & 0.44 & 0.40  &  -5,(-4) \nl
3C78        &   cD              &  S0/a  & 0.77 & 0.49 & 0.42  &  0 \nl
3C88        &   cD              &  E pec? & 0.80 & 0.49 & 0.42  &  -5,(-4) \nl
3C98        &   cD              &  E1?    & 0.85 & 0.47 & 0.41  &  -5,(-4) \nl
3C192       &   cD              &  XXX    & 0.83 & 0.38 & 0.32  &  17,(-4) \nl
3C264       &   cD              &  E      & 0.74 & 0.41 & 0.35  &  -5,(-4) \nl
3C293       &   cD              &  S?     & 0.77 & 0.30 & 0.40  &  1-7,(-4) \nl
3C296       &   cD              &  XXX    & 0.88 & 0.45 & 0.35  &  17 \nl
3C305       &   cD              &  SB0    & 0.85 & 0.23 & 0.25  &  -2 \nl
Cl1604      &   G1              &  XXX    & 0.80 & 0.27 & 0.36  &  17 \nl
PHL1093     &   G2              &  XXX    & 0.59 & 0.31 & 0.26  &  17 \nl
Cl1318      &   G1              &  XXX    & 0.75 & 0.36 & 0.43  &  17 \nl
Cl1612      &   G1              &  XXX    & 0.60 & 0.21 & 0.31  &  17 \nl
Cl1610      &   G1              &  XXX    & 0.52 & 0.33 & 0.26  &  17 \nl
Cl0948      &   ??              &  XXX    & 0.81 & 0.33 & 0.41  &  17 \nl
Cl1049      &   G1              &  XXX    & 0.72 & 0.49 & 0.27  &  17 \nl
Cl1607      &   G1              &  XXX    & 0.64 & 0.50 & 0.27  &  17 \nl
Cl0948.9    &   ??              &  XXX    & 0.63 & 0.31 & 0.45  &  17 \nl
Cl1446a     &   G3              &  XXX    & 0.70 & 0.35 & 0.26  &  17 \nl
Cl1446b     &   G4              &  XXX    & 0.92 & 0.14 & 0.23  &  17 \nl
Cl0949      &   ??              &  XXX    & 0.75 & 0.42 & 0.23  &  17 \nl
Cl0024      &   G1              &  XXX    & 0.87 & 0.38 & 0.45  &  17 \nl
CRBCL       &   G2              &  XXX    & 0.95 & 0.45 & 0.39  &  17 \nl
Cl1534      &   G1              &  XXX    & 0.81 & 0.42 & 0.39  &  17 \nl
A665        &   G1              &  XXX    & 0.83 & 0.42 & 0.43  &  17 \nl
HYG8        &   G9              &  XXX    & 0.82 & 0.35 & 0.36  &  17 \nl
A2317       &   G1              &  XXX    & 0.77 & 0.38 & 0.39  &  17 \nl
Cl1446c     &   G1              &  XXX    & 0.56 & 0.04 & 0.27  &  17 \nl
A1961       &   G1              &  XXX    & 0.74 & 0.36 & 0.32  &  17 \nl
Cl0025      &   G1              &  XXX    & 0.75 & 0.25 & 0.37  &  17 \nl
3C28        &   ??              &  XXX    & 0.53 & 0.31 & 0.35  &  17 \nl
3C219       &   ??              &  XXX    & 0.63 & 0.28 & 0.29  &  17 \nl
3C198       &   pec             &  XXX    & 0.34 & 0.06 & 0.12  &  17 \nl
3C277       &   pec             &  XXX    & 0.67 & 0.32 & 0.31  &  17 \nl
Mrk6        &   Sy1             &  SAB0+: Sy1.5 & 0.11 & 0.15 & 0.25  &  -2 \nl
Mrk9        &   Sy1             &  S0 pec? Sy1 & -0.34 & -0.08 & 0.02  & -2 \nl
Mrk10       &   Sy1             &  SBbc Sy1 & -0.36 & -0.02 & 0.08  &  4 \nl
Mrk42       &   Sy1             &  SBb Sy1  & 0.05 & 0.04 & 0.11  &  3 \nl
Mrk50       &   Sy1             &  E? Sy1   & -0.28 & 0.01 & 0.10  &  -5 \nl
Mrk69       &   Sy1             &  Sy1      & -0.15 & -0.17 & -0.02  &  17 \nl
Mrk141      &   Sy1             &  E Sy1.5  & -0.07 & -0.01 & 0.10  &  -5 \nl
Mrk142      &   Sy1             &  S? Sy1   & -0.31 & -0.13 & -0.02  &  1-7 \nl
Mrk279      &   Sy1             &  S0 Sy1.5 & -0.45 & -0.17 & 0.02  &  -2      
\nl
Mrk290      &   Sy1             &  E1? Sy1     & -0.54 & -0.13 & -0.13  &  -5 
\nl
Mrk291      &   Sy1             &  SB(s)a: Sy1 & -0.08 & -0.11 & 0.08  &  1 \nl
Mrk304      &   Sy1             &  compact Sy1 & -0.52 & -0.16 & -0.15  &  11,-6 
\nl
Mrk315      &   Sy1             &  E1 pec? Sy1.5 & 0.22 & 0.07 & 0.21  & -5 \nl
Mrk335      &   Sy1             &  S0/a Sy1      & -0.60 & -0.20 & -0.21  & 0 
\nl
Mrk352      &   Sy1             &  SA0 Sy1       & -0.47 & -0.09 & -0.03  & -2 
\nl
Mrk358      &   Sy1             &  SAB(rs)bc: Sy1 & 0.02 & 0.12 & 0.18  & 4 \nl
Mrk464      &   Sy1             &  compact Sy1.5  & -0.20 & -0.07 & 0.05  & 
11,-6 \nl
Mrk474      &   Sy1             &  SB(s)0/a? Sy1  & 0.01 & 0.21 & 0.15  & 0 \nl
Mrk478      &   Sy1             &  compact Sy1    & -0.40 & -0.22 & -0.15  & 
11,-6 \nl
Mrk486      &   Sy1             &  SBb? Sy1       & -0.34 & -0.11 & -0.10  & 3 
\nl
Mrk506      &   Sy1             &  SAB(r)a,Sy1.5  & -0.03 & 0.24 & 0.20  & 1 \nl
Mrk79       &   Sy1-merger      &  SBb Sy1.2      & -0.51 & -0.11 & -0.06  & 3 
\nl
Mrk110      &   Sy1-merger      &  pair? Sy1      & -0.44 & -0.20 &-0.07  & 17 
\nl
Mrk231      &   Sy1-merger      &  SA(rs)c? pec Sy1 & 0.51 & 0.11 & 0.25  & 5 
\nl
Mrk618      &   Sy1             &  SB(s)b pec Sy1 & -0.35 & -0.12 & -0.05  & 3 
\nl
IIZw136     &   Sy1             &  S? Sy1         & -0.45 & -0.24 & -0.20  & 1-7 
\nl
3C390.3     &   Sy1             &  opt.var. Sy1   & -0.40 & -0.04 & -0.01  & 17 
\nl
PKS2349-01  &   Sy1             &  XXX            & -0.48 & -0.23 & -0.13  & 17 
\nl
Mrk1        &   Sy2             &  (R')S? Sy2     & 0.39 & 0.10 & 0.20  & 1-7 
\nl
Mrk34       &   Sy2             &  S Sy2          & 0.40 & 0.15 & 0.16  & 1-7 
\nl
Mrk176      &   Sy2             &  SA(s)0/a pec: Sy2 (triplet) & 0.76
& 0.24 & 0.26  & 0 \nl
Mrk268      &   Sy2             &  SBb Sy2 pair? & 0.70 & 0.21 & 0.32  & 3 \nl
Mrk348      &   Sy2             &  SA(s)0/a: Sy2 & 0.49 & 0.22 & 0.30  & 0 \nl
Mrk372      &   Sy2             &  S0/a Sy1.5    & 0.68 & 0.35 & 0.37  & 0 \nl
Mrk78       &   Sy2-merger      &  SB Sy2        & 0.71 & 0.18 & 0.30  & 1-7 \nl
Mrk463      &   Sy2-merger      &  triplet=Sy2+Sy1+? & 0.01 & -0.02 & 0.04  & 17 
\nl
IRAS-F17242+6339 & AGN          &  Sy2  & 0.34 & 0.38 & 0.35  & 17 \nl
IRAS-F17418+7042 & AGN          &  Sy2  & 0.13 & -0.13 & 0.07  & 17 \nl
IRAS-F17428+6251 & AGN          &  Sy2  & 0.11 & 0.03 & 0.54  & 17 \nl
IRAS-F17226+6844 & AGN-burst    &  Irr  & -0.18 & -0.30 & -0.06   & 17 \nl
IRAS-F17462+6402 & AGN-burst    &  XXX  & 0.36 & 0.18 & 0.50  & 17 \nl
IRAS-F18066+6632 & AGN-burst    &  XXX  & -0.26 & -0.04 & 0.34   & 17 \nl
IRAS-F18173+6617 & AGN-burst    &  XXX  & 0.35 & 0.04 & 0.44  & 17 \nl
IRAS-17297+6900  & AGN-burst    &  XXX  & 0.01 & -0.15 & 0.27   & 17 \nl
IRAS-18121+6256  & AGN-burst    &  XXX  & 0.11 & 0.00 & 0.19  & 17 \nl
IRAS-18423+6717  & AGN-burst    &  XXX  & 0.52 & 0.25 & 0.77   & 17 \tablebreak
IRAS-F17242+6637 & burst        &  XXX  & 0.34 & 0.38 & 0.35   & 17 \nl
IRAS-F17425+6615 & burst        &  XXX  & 0.13 & -0.13 & 0.07   & 17 \nl
IRAS-18234+6440  & burst        &  XXX  & -0.06 & -0.19 & 0.12  & 17 \nl
IRAS-F17330+7017 & burst        &  XXX  & 0.26 & 0.31 & 0.32 & 17 \nl
IRAS-F18031+6312 & burst        &  XXX  & 0.02 & -0.23 & -0.02   & 17 \nl
IRAS-F18014+6318 & burst        &  XXX  & 0.39 & -0.16 & 0.27  & 17 \nl
\enddata
\tablecomments{The Rakos sample does not meet the large aperture criterion. It 
is rather a compilation of a wide range of spectral types found in Kennicutt 
(1992a), Ashby, Houck \& Hacking (1992), Gunn \& Oke (1975), Yee \& Oke (1978) 
and de Bruyn \& Sargent (1978). This serves here to asses the necessity for the 
large aperture criterion, and to test whether the Kinney + Kennicutt sample 
covers the entire parameter space occupied by galaxies. \\
N.B.: In the Rakos classification (column 2), GX - where X is a number - denotes 
the Xth brightest galaxy of the cluster listed in column 1. Galaxies with 
morphologies so peculiar that no Hubble type can be assigned are given T=15 and 
galaxies for which no morphological classification could be found are listed as 
T=17.}
\end{deluxetable}

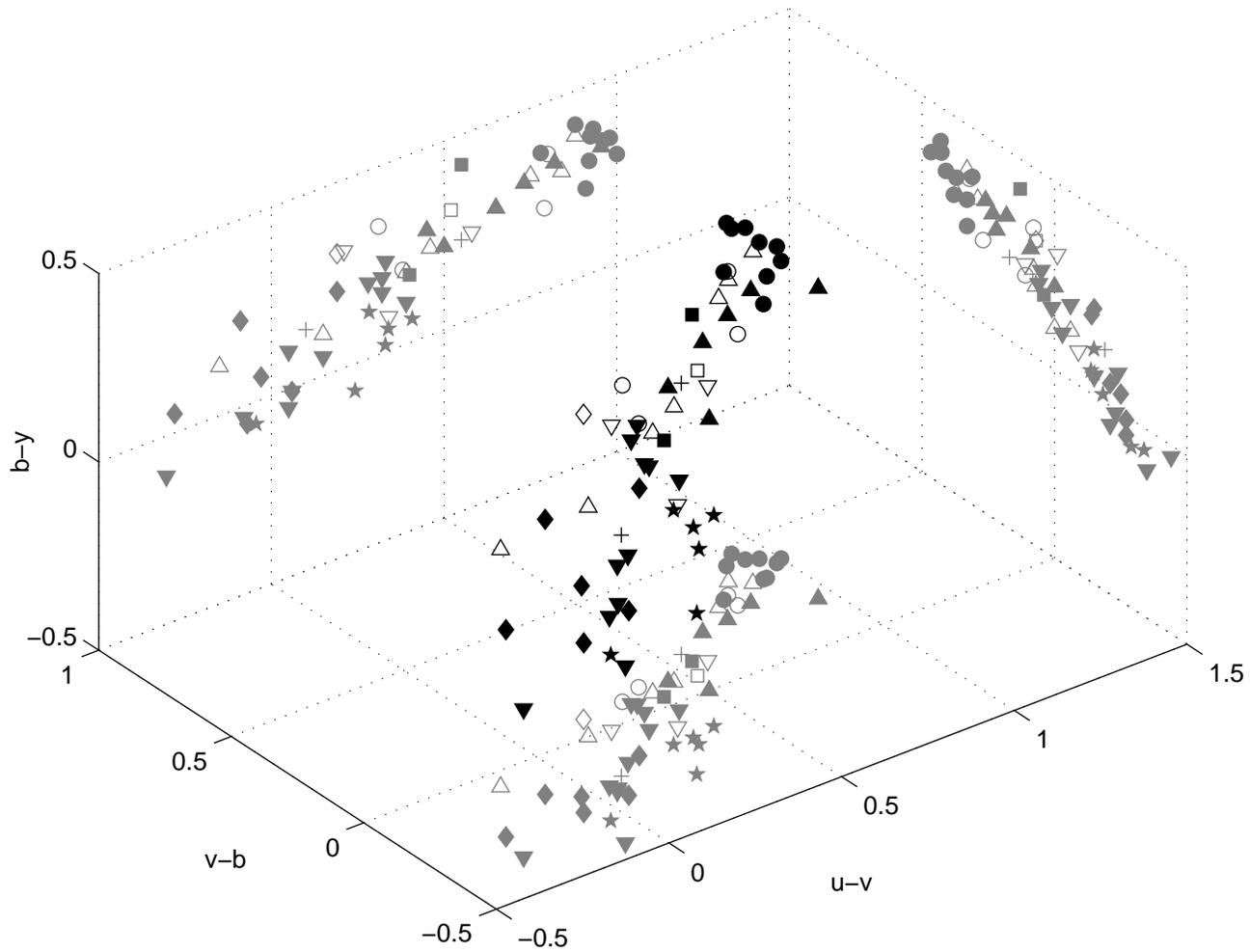
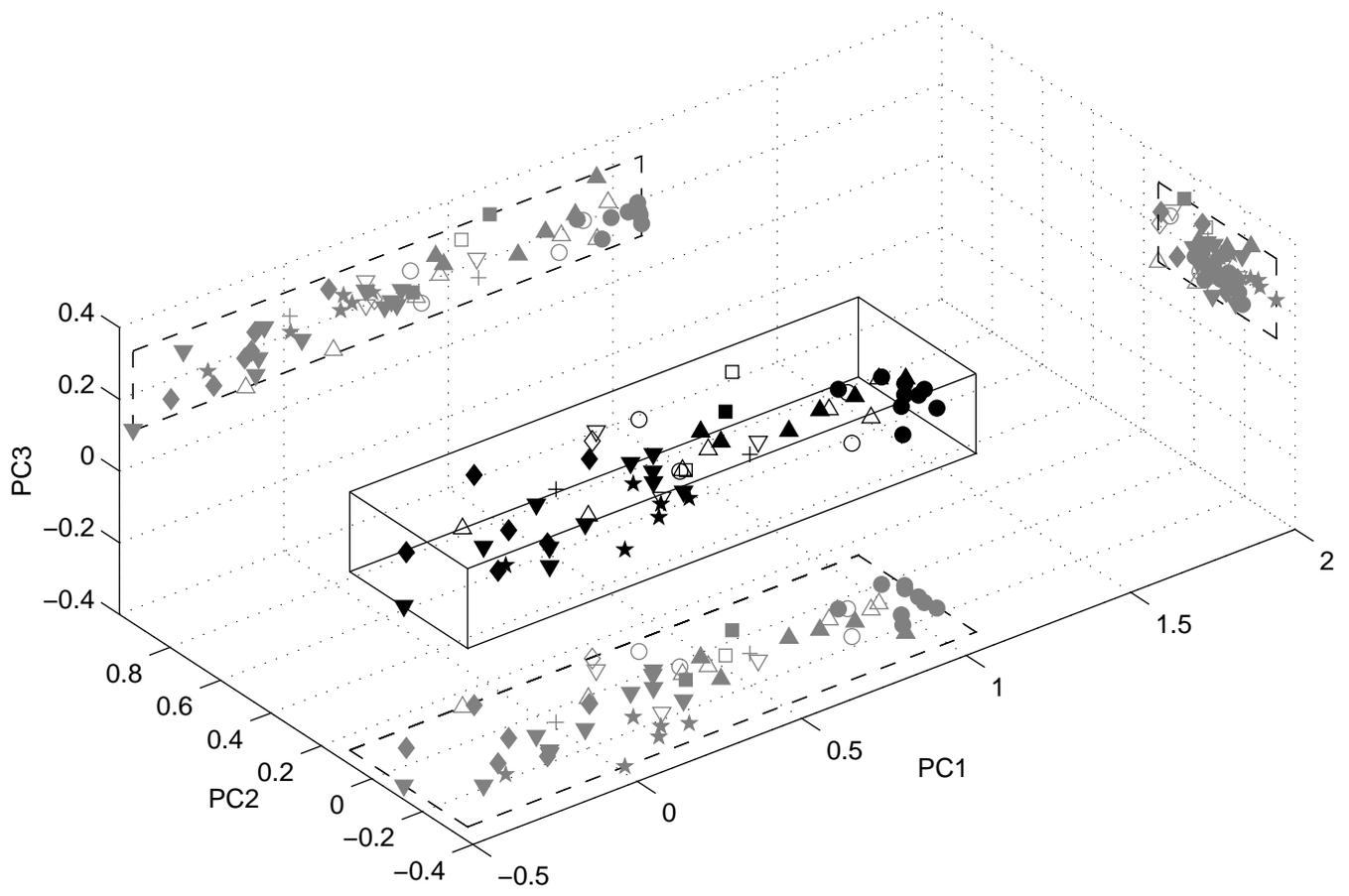
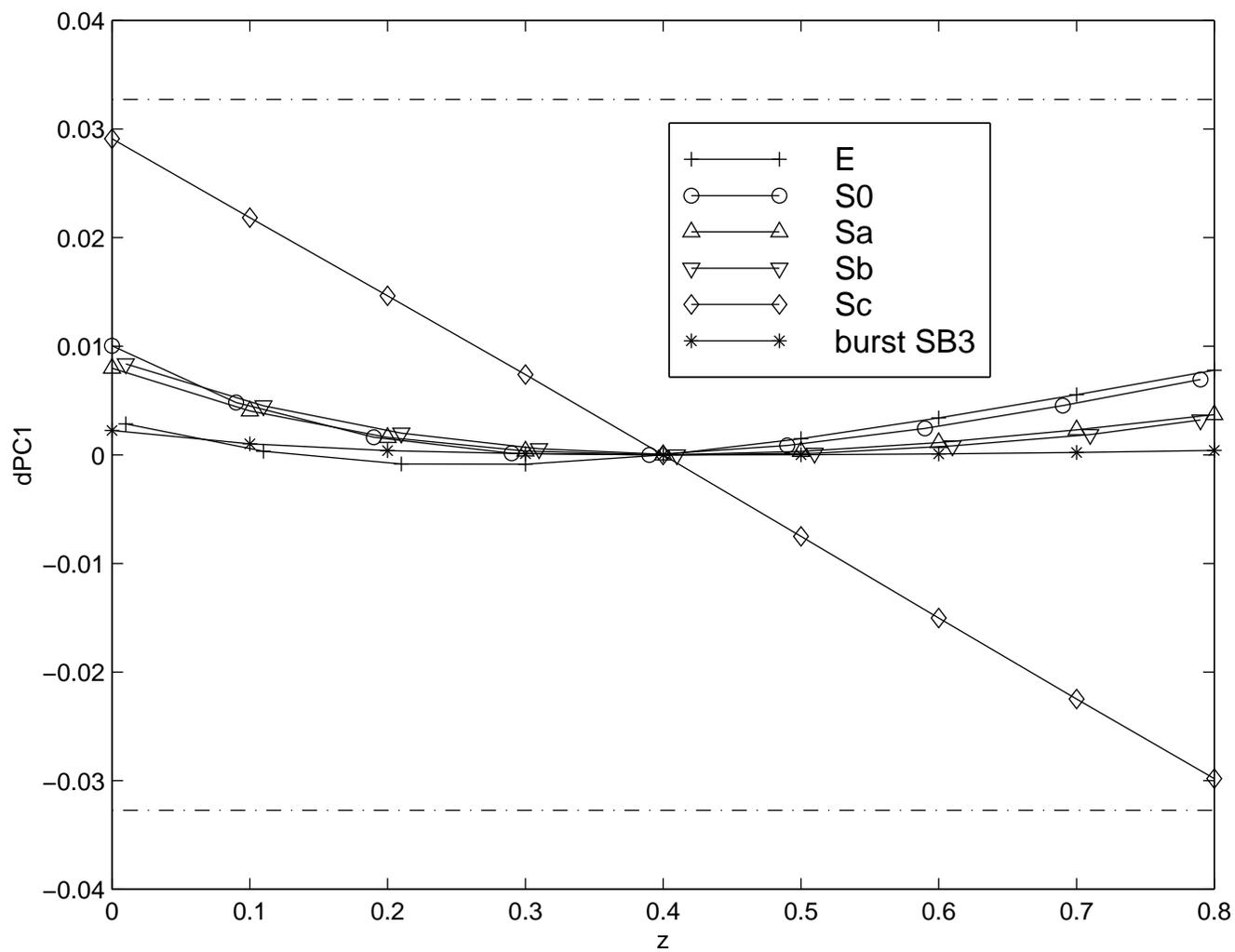
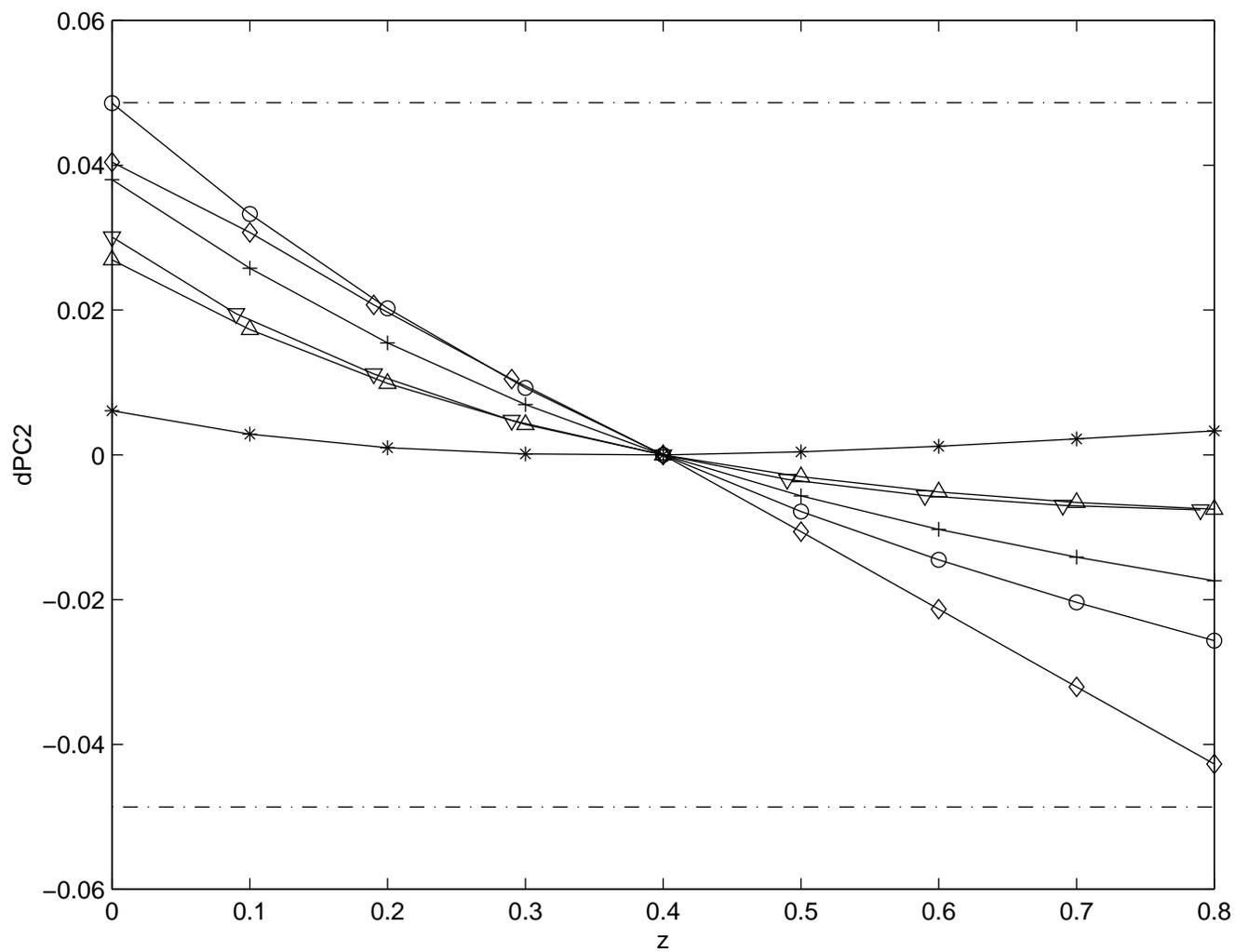
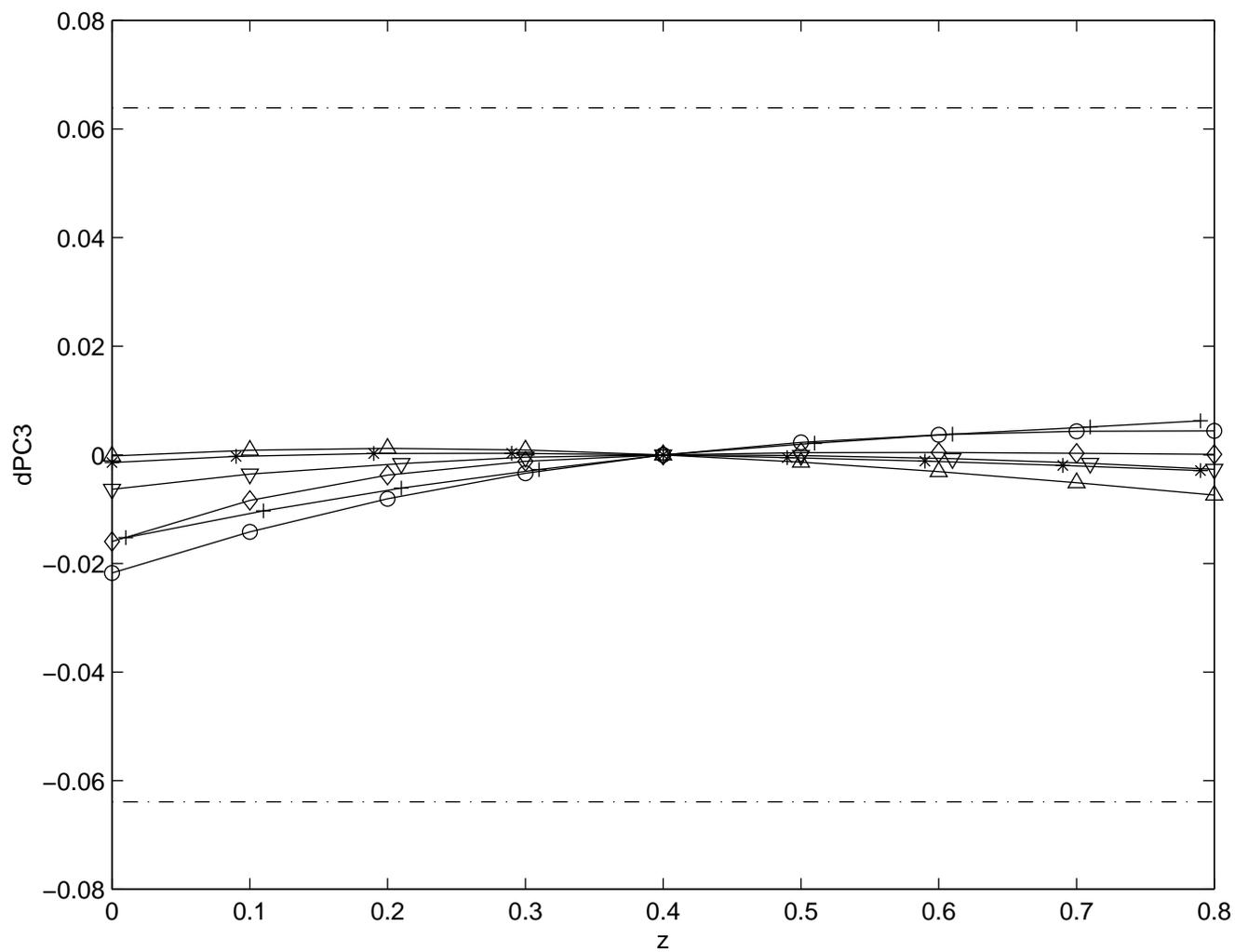
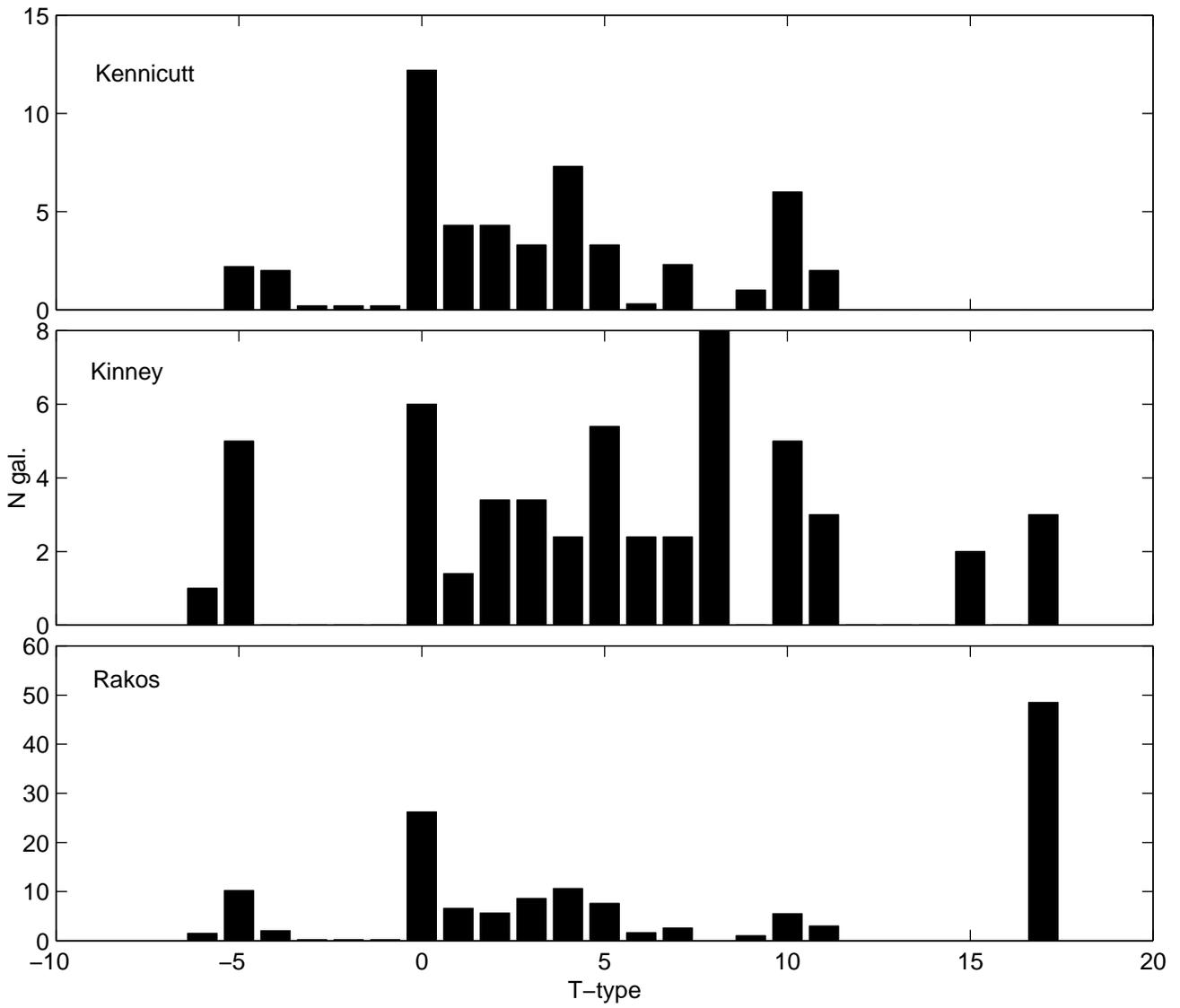
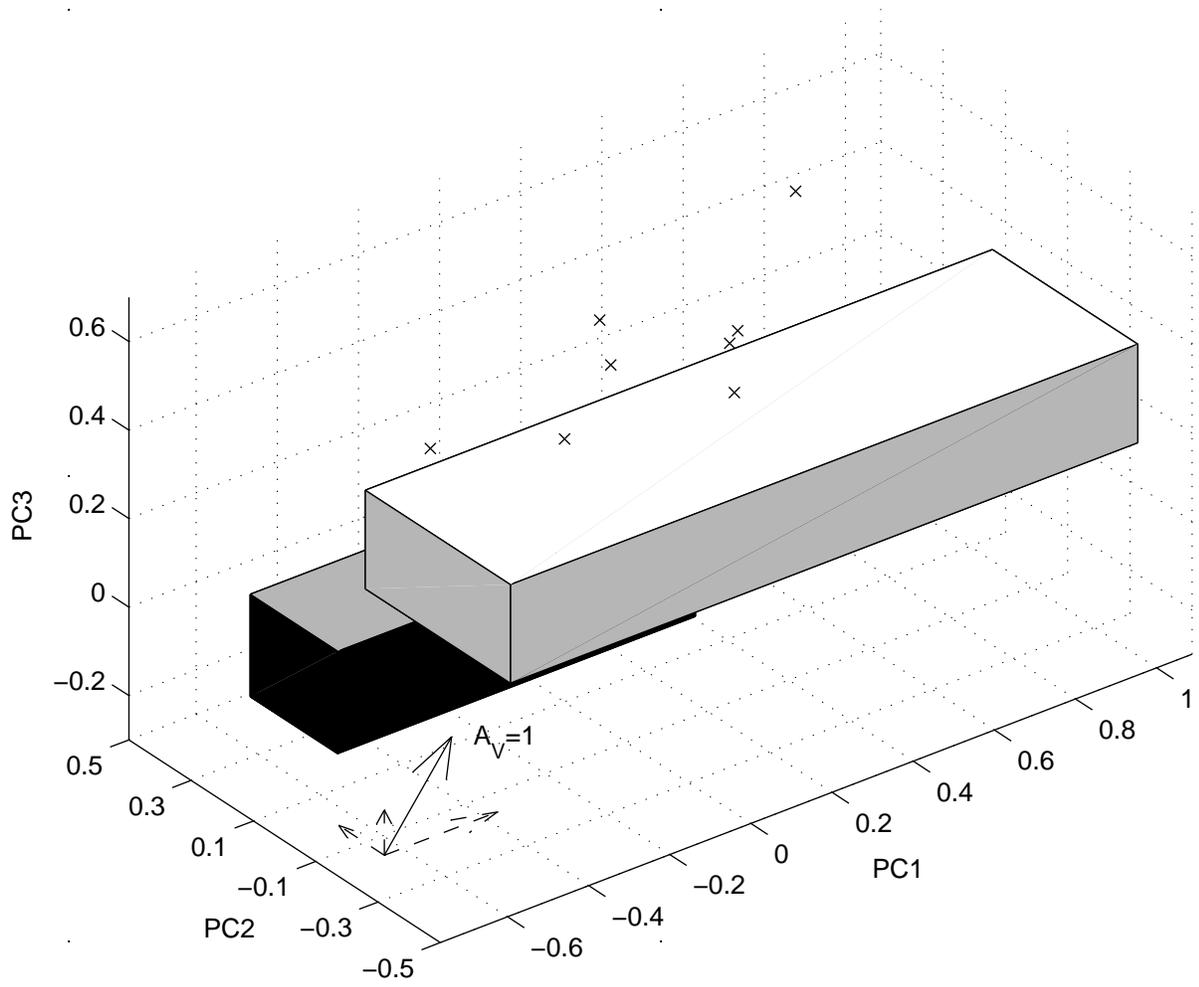
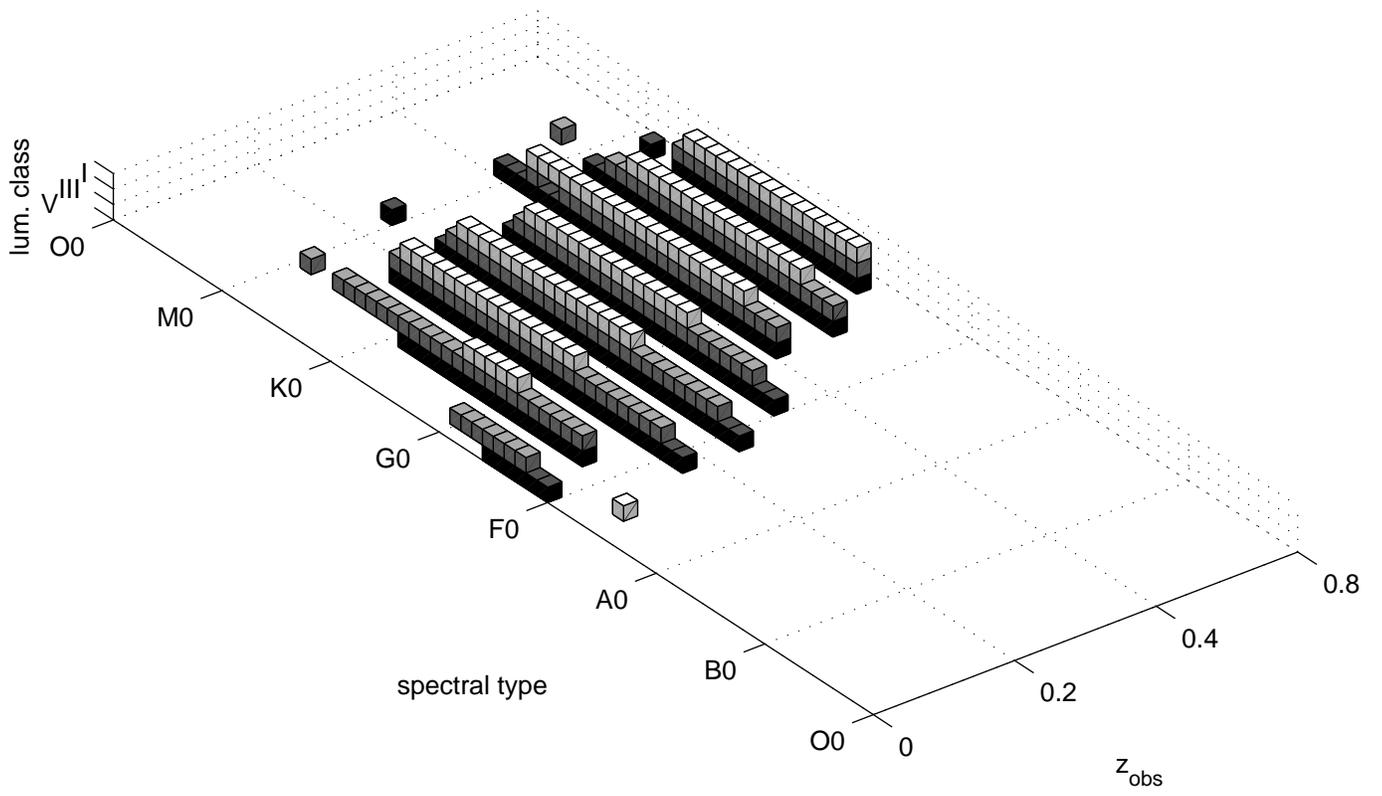
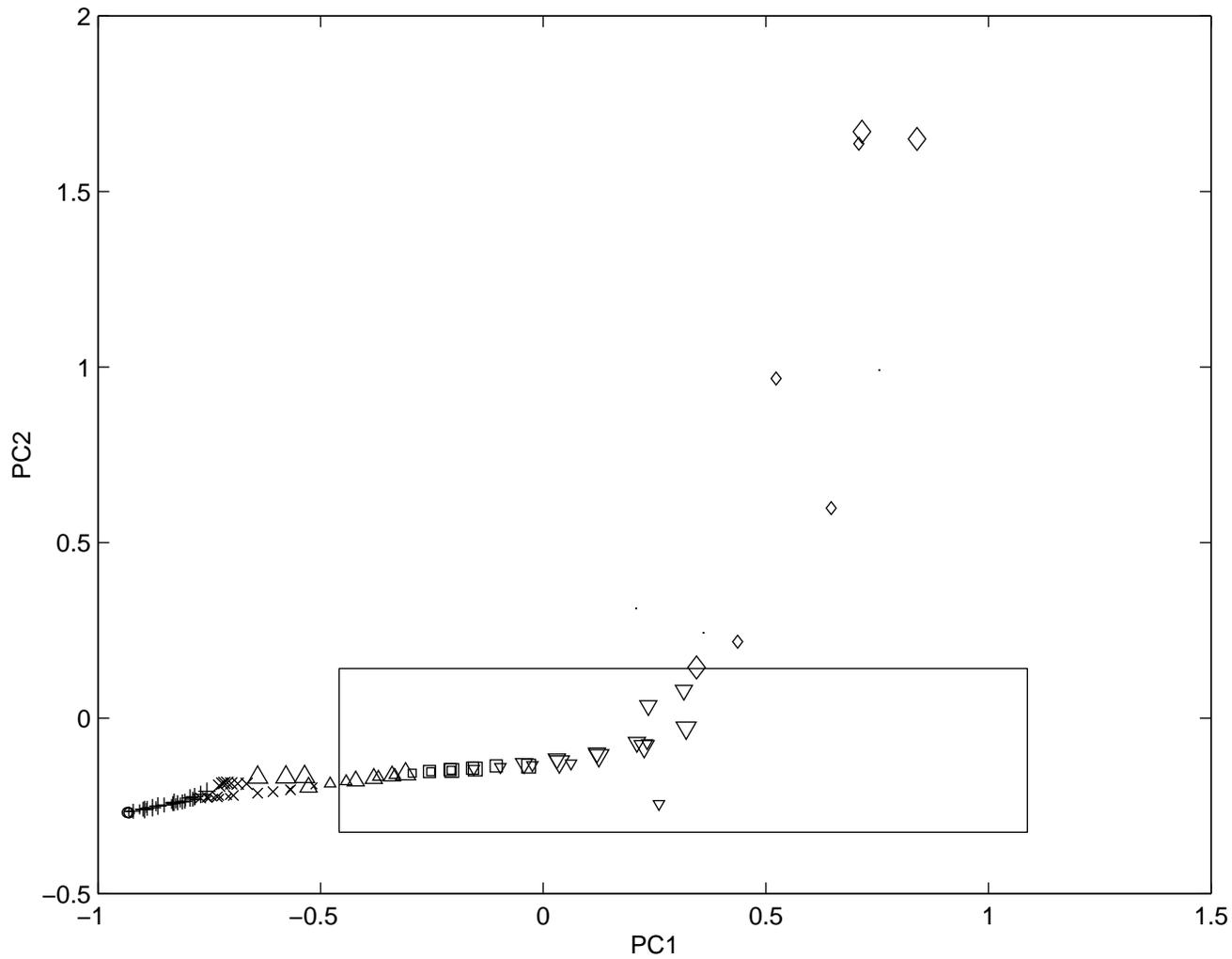
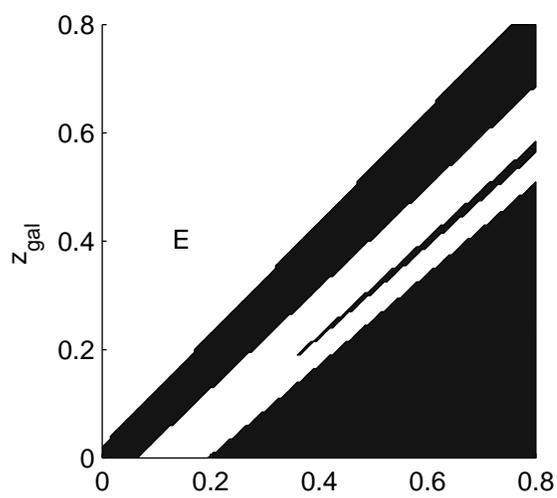
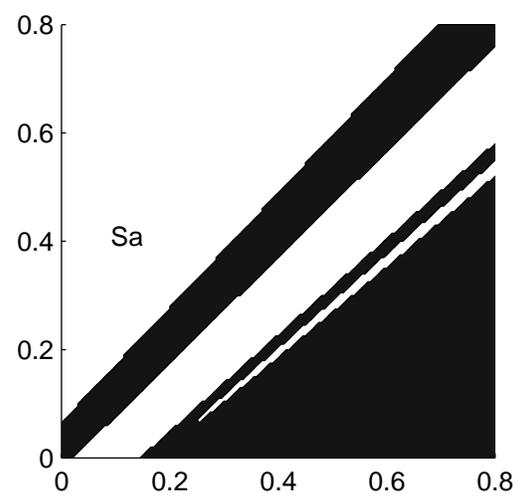
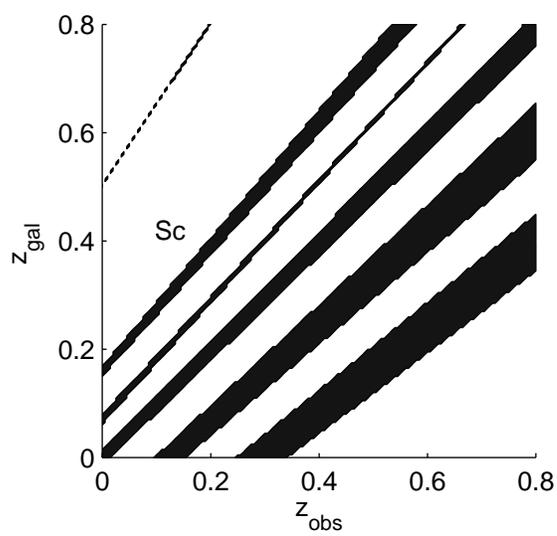
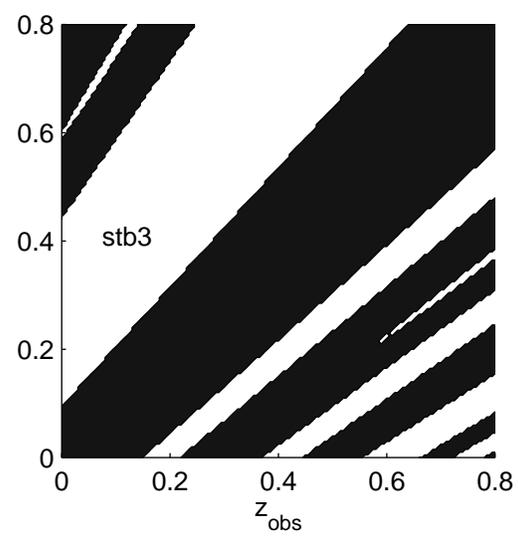
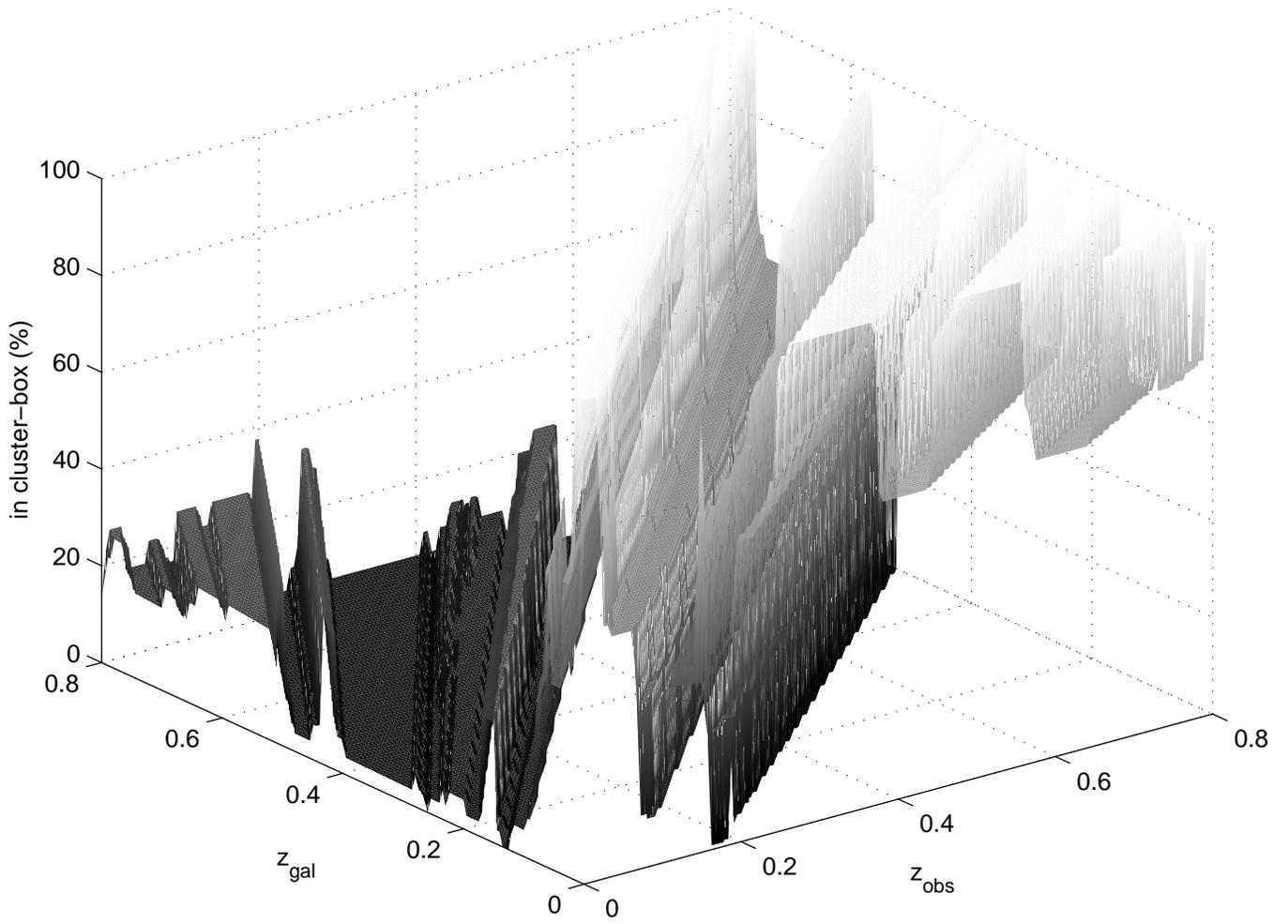
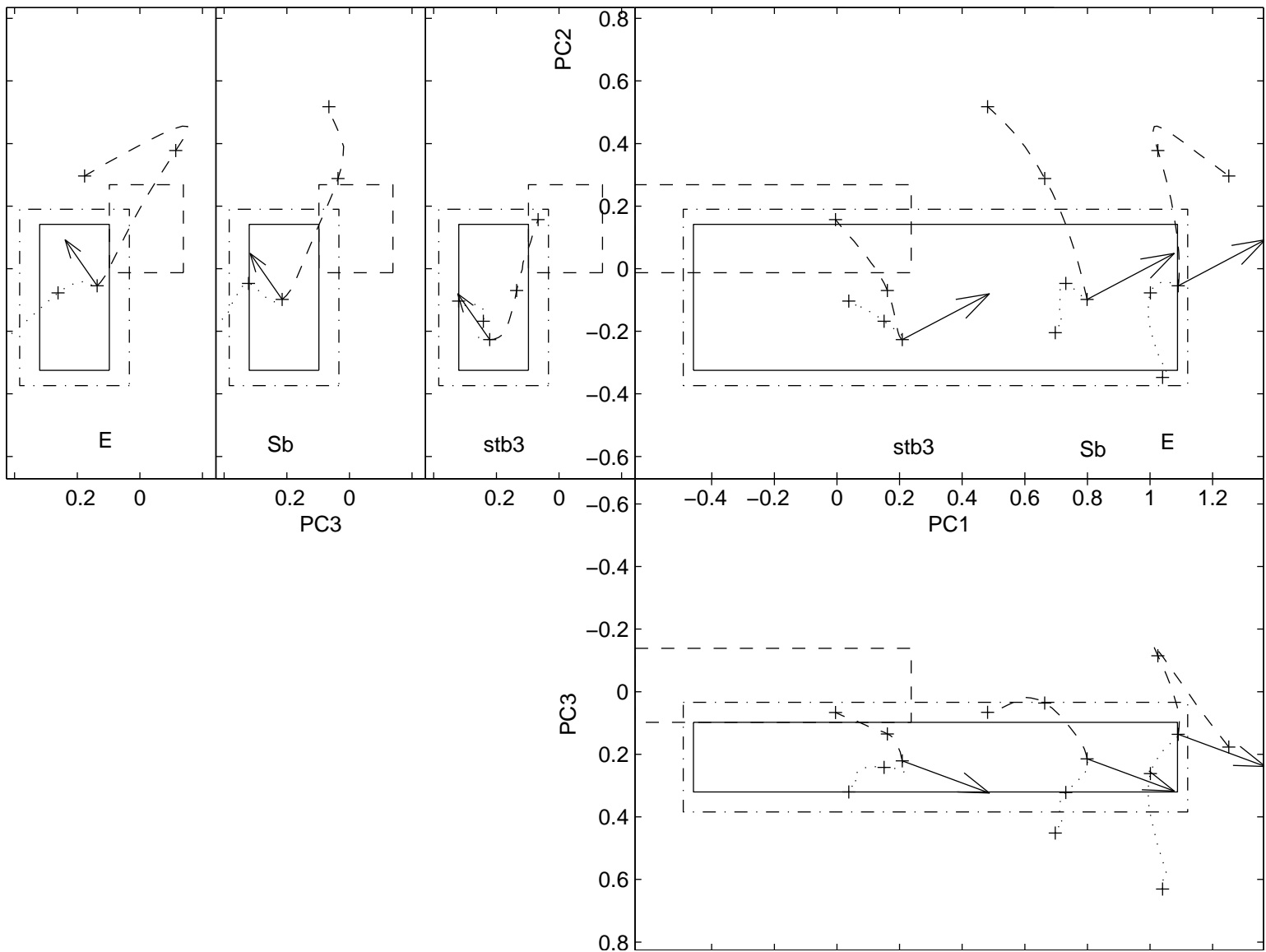
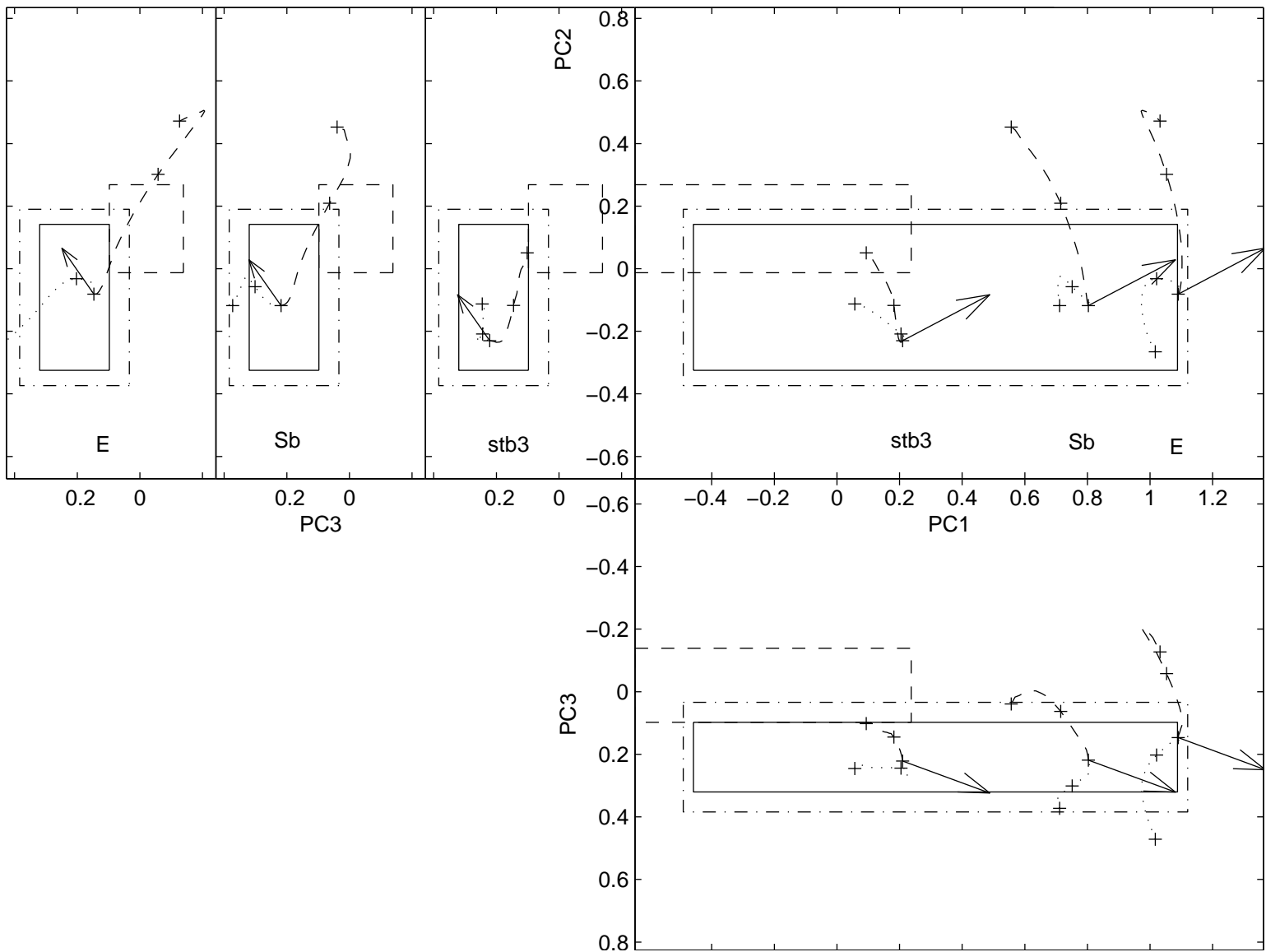
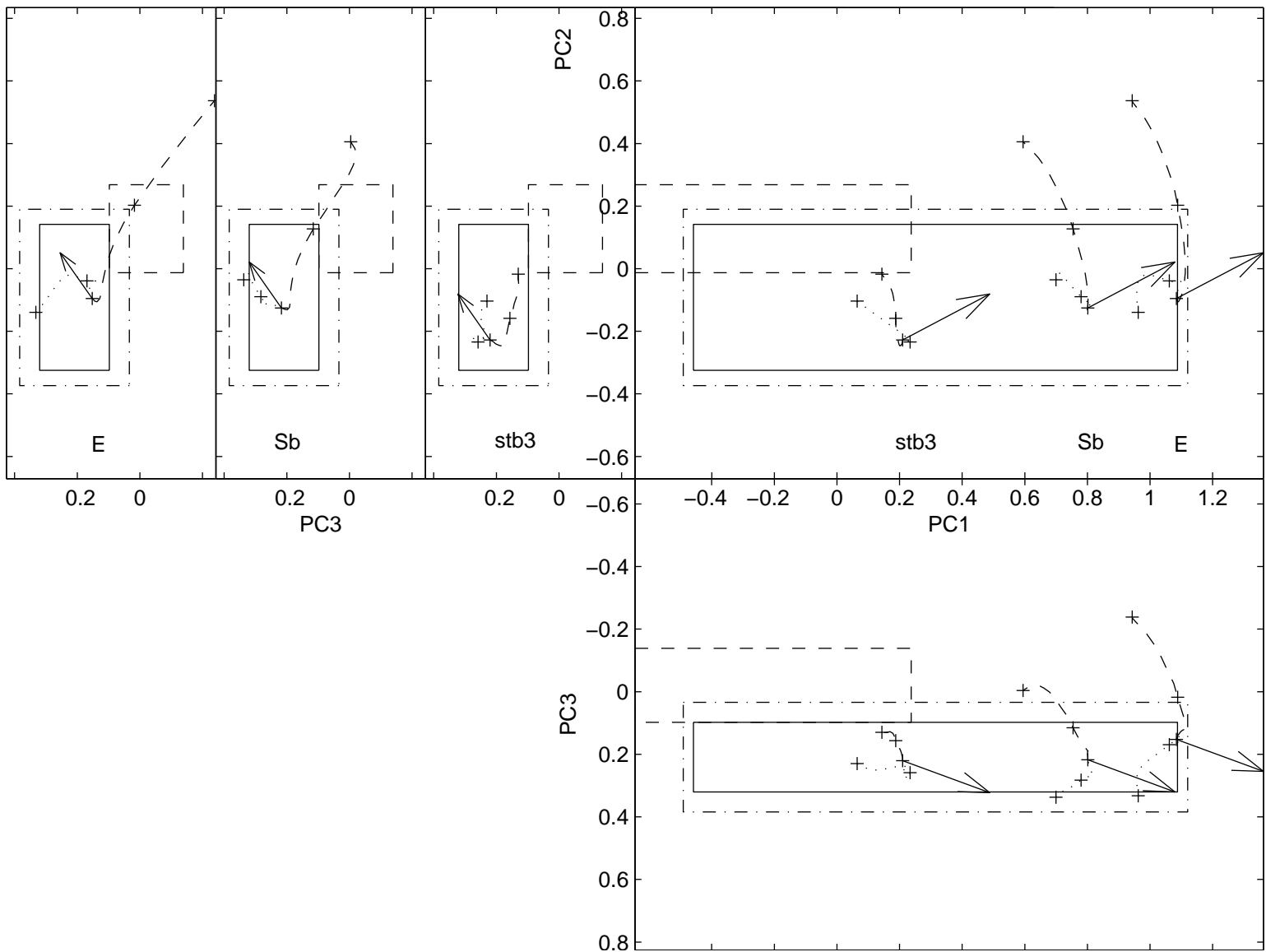
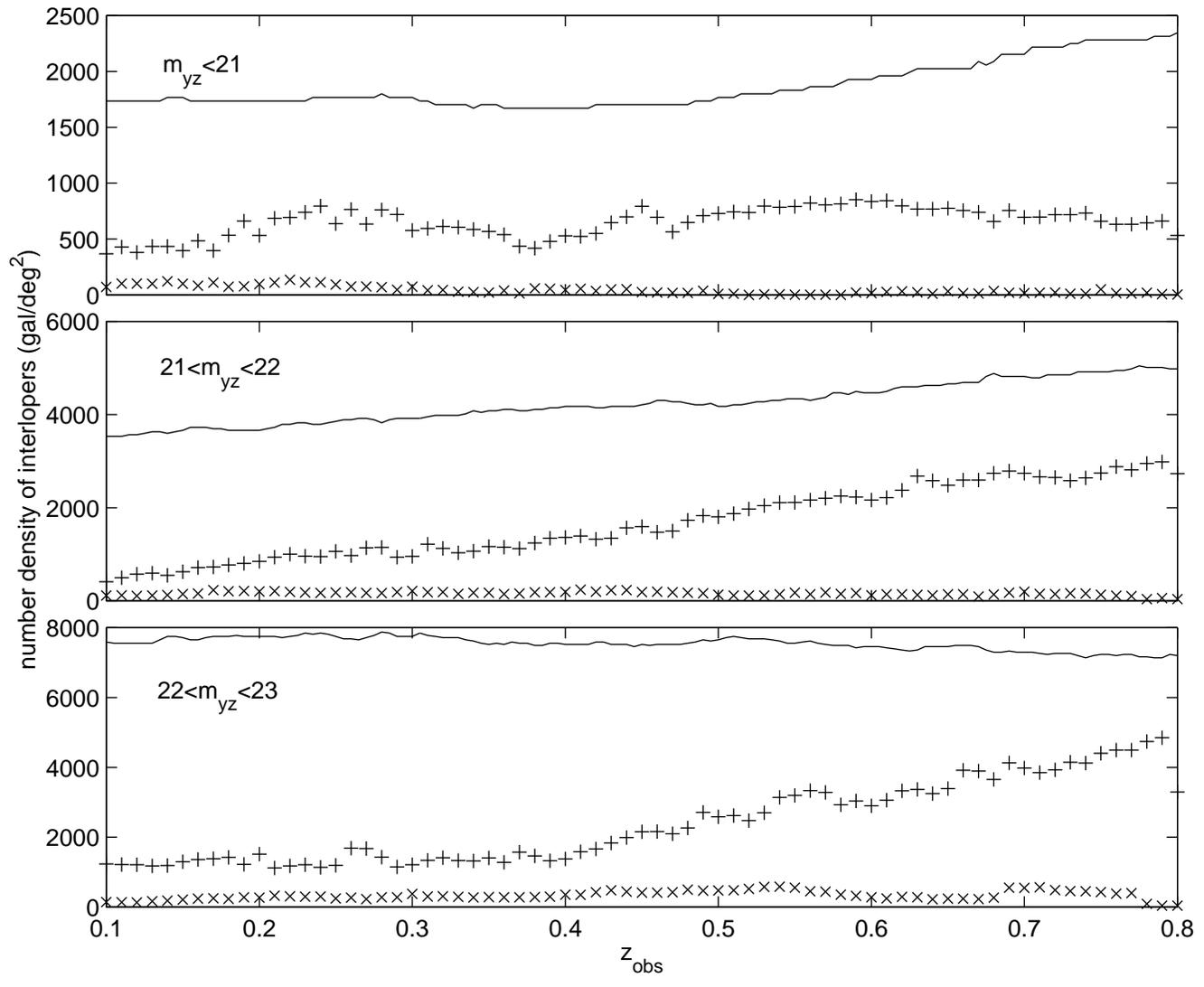
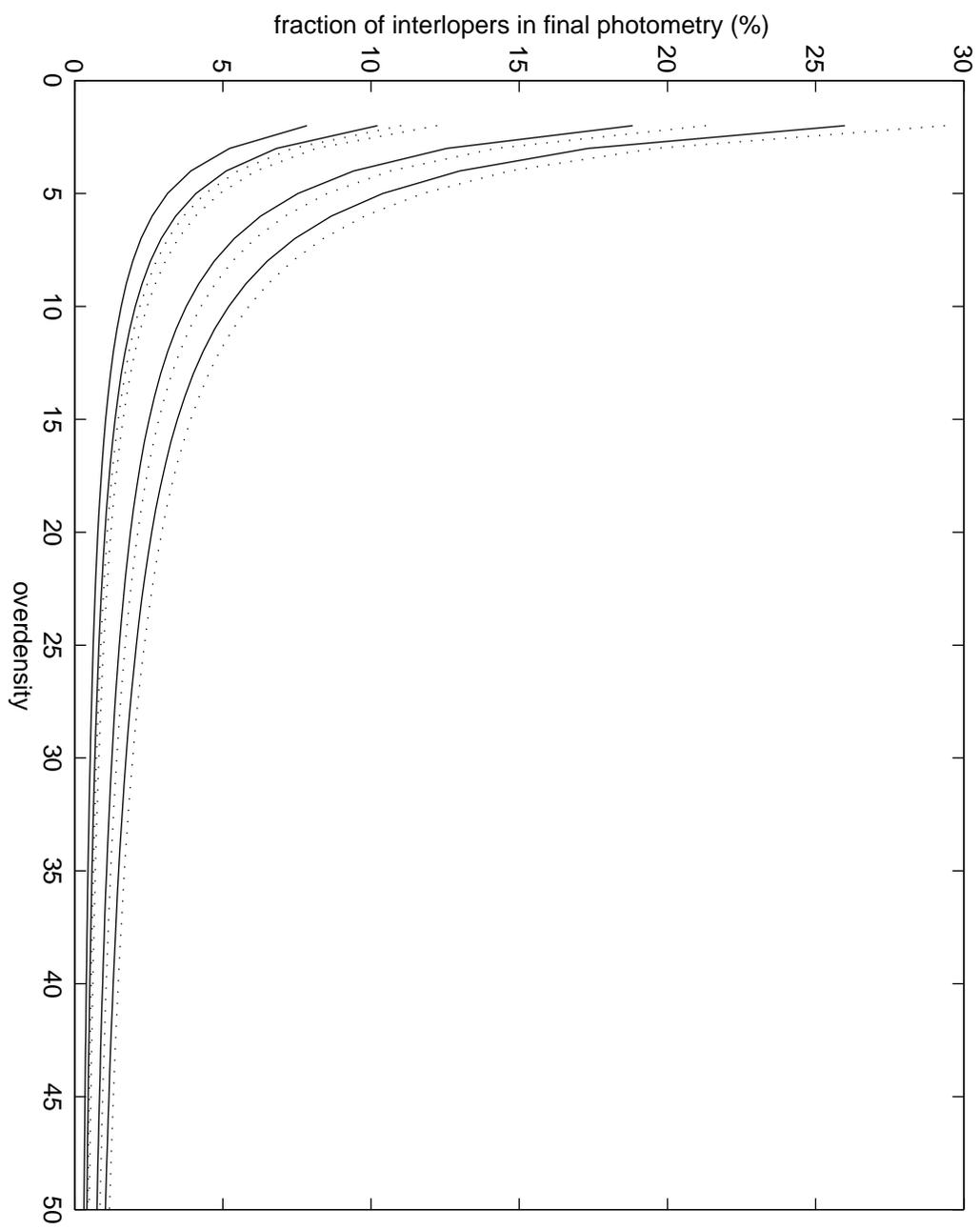
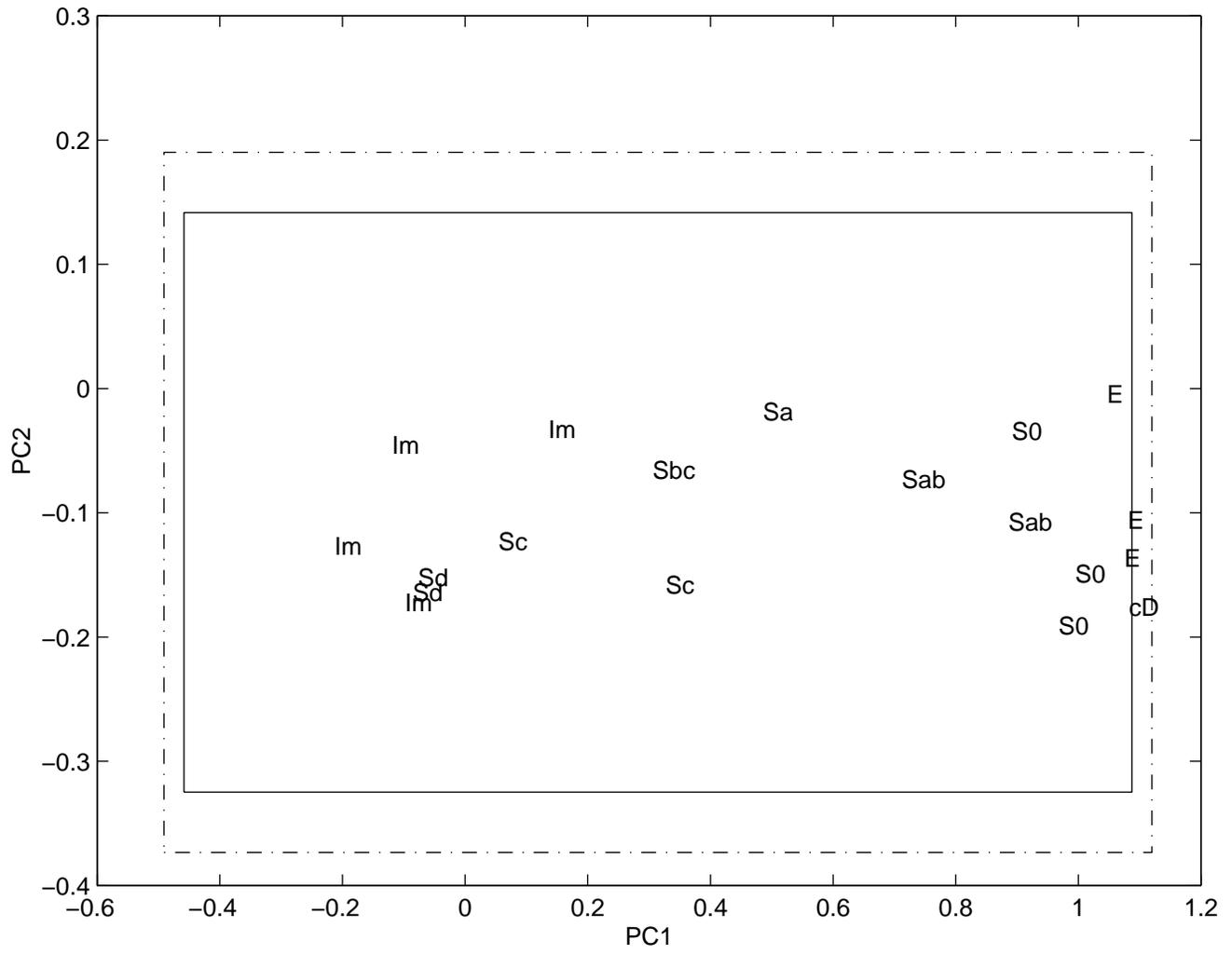
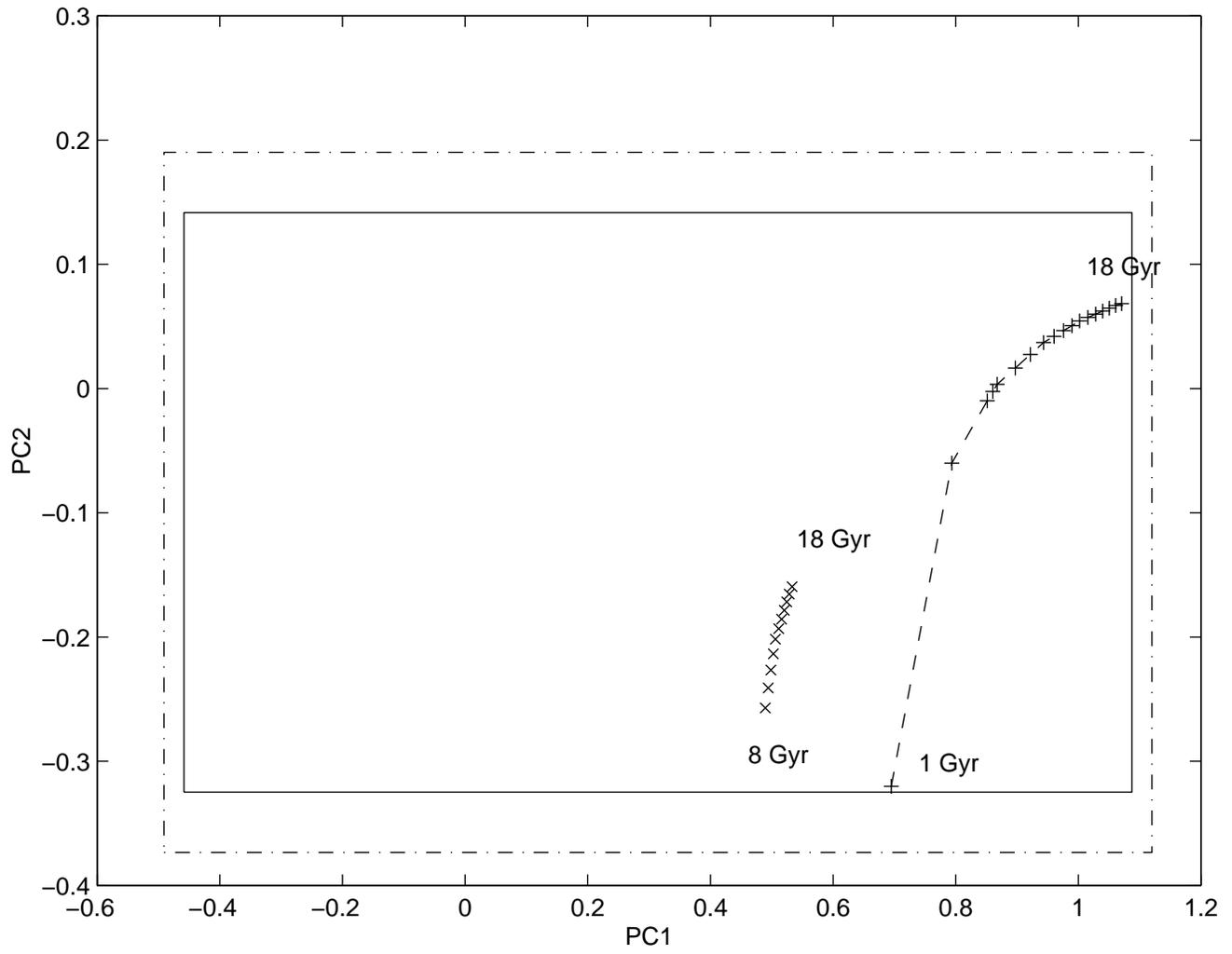
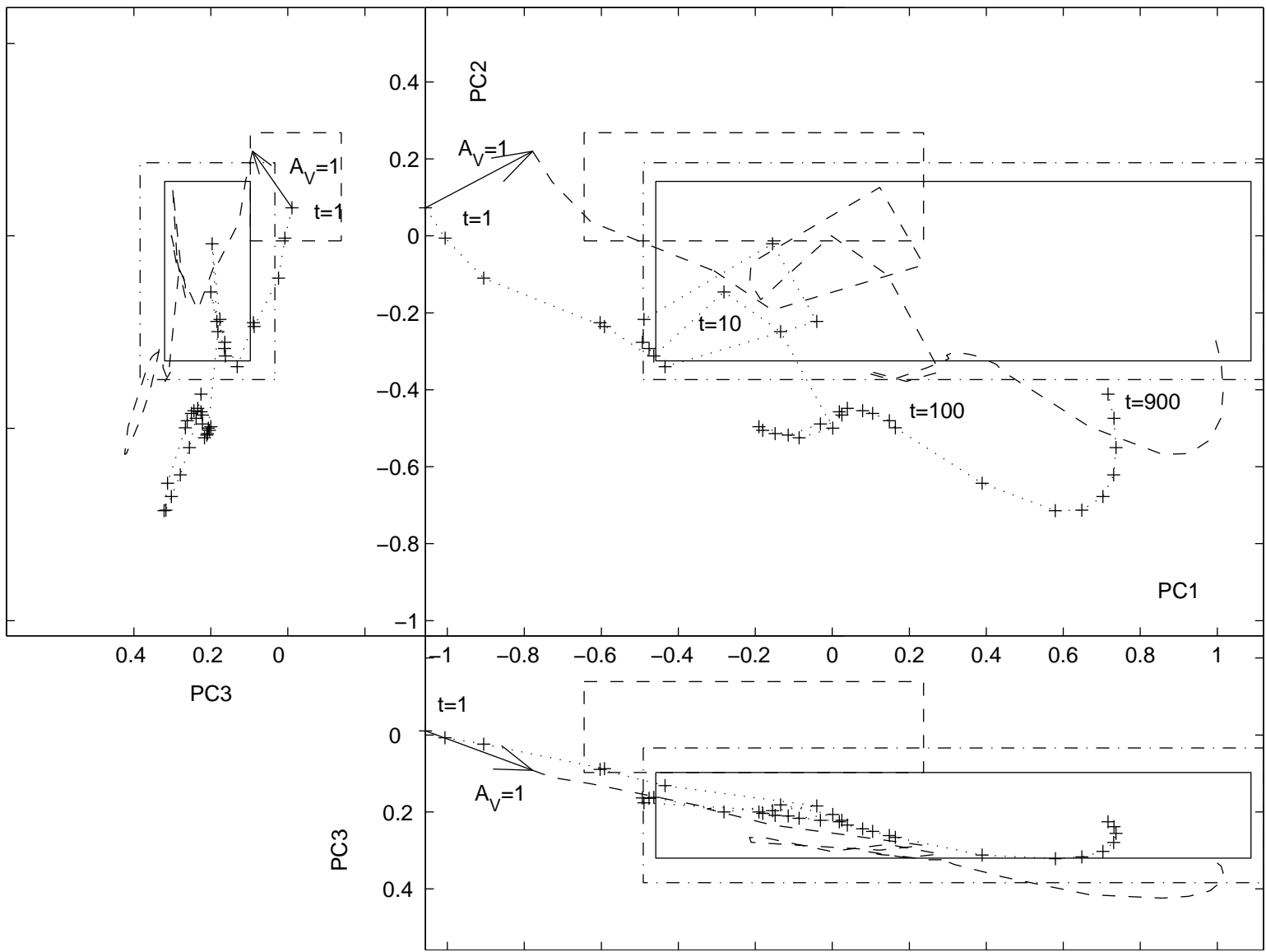
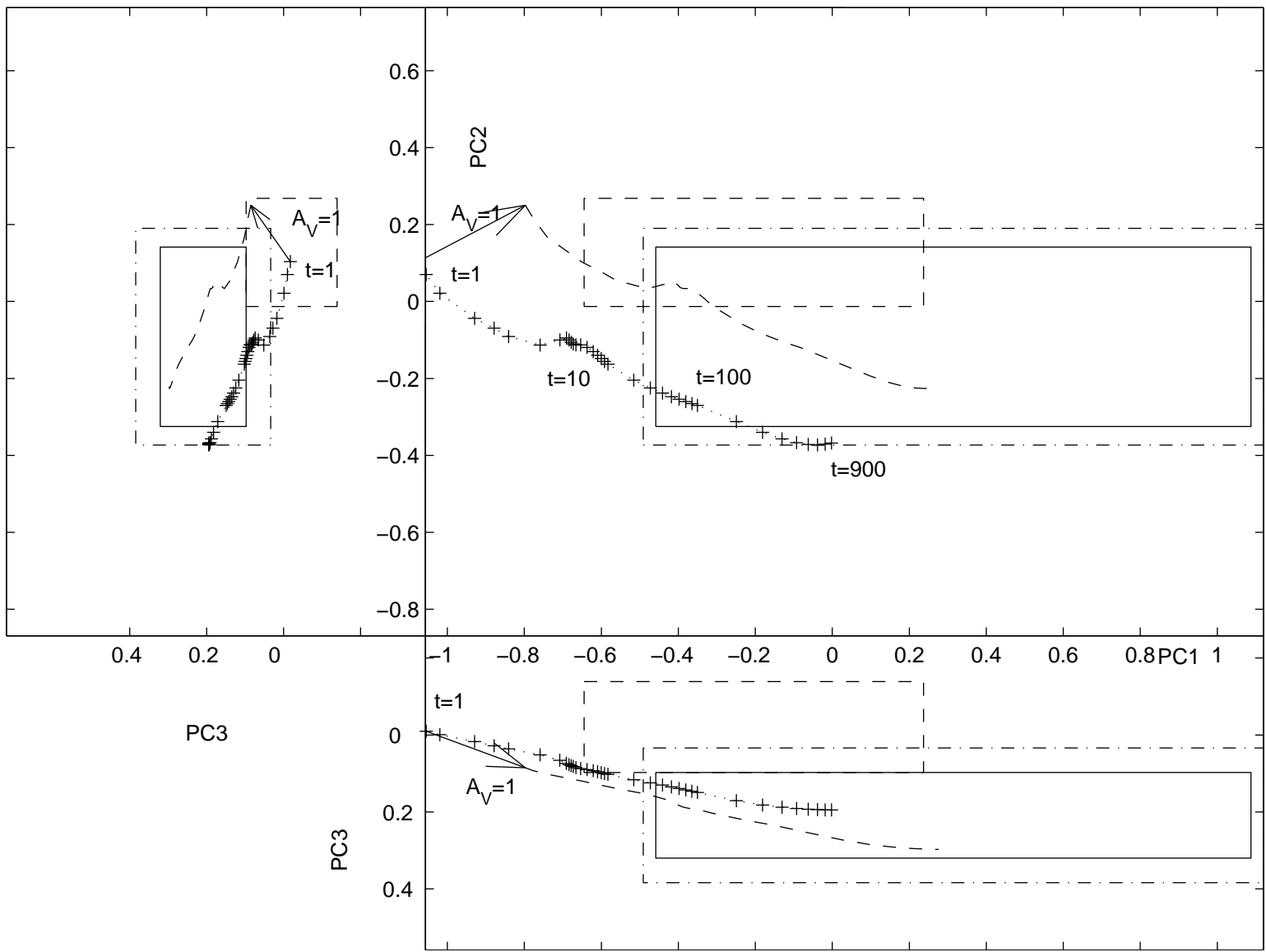
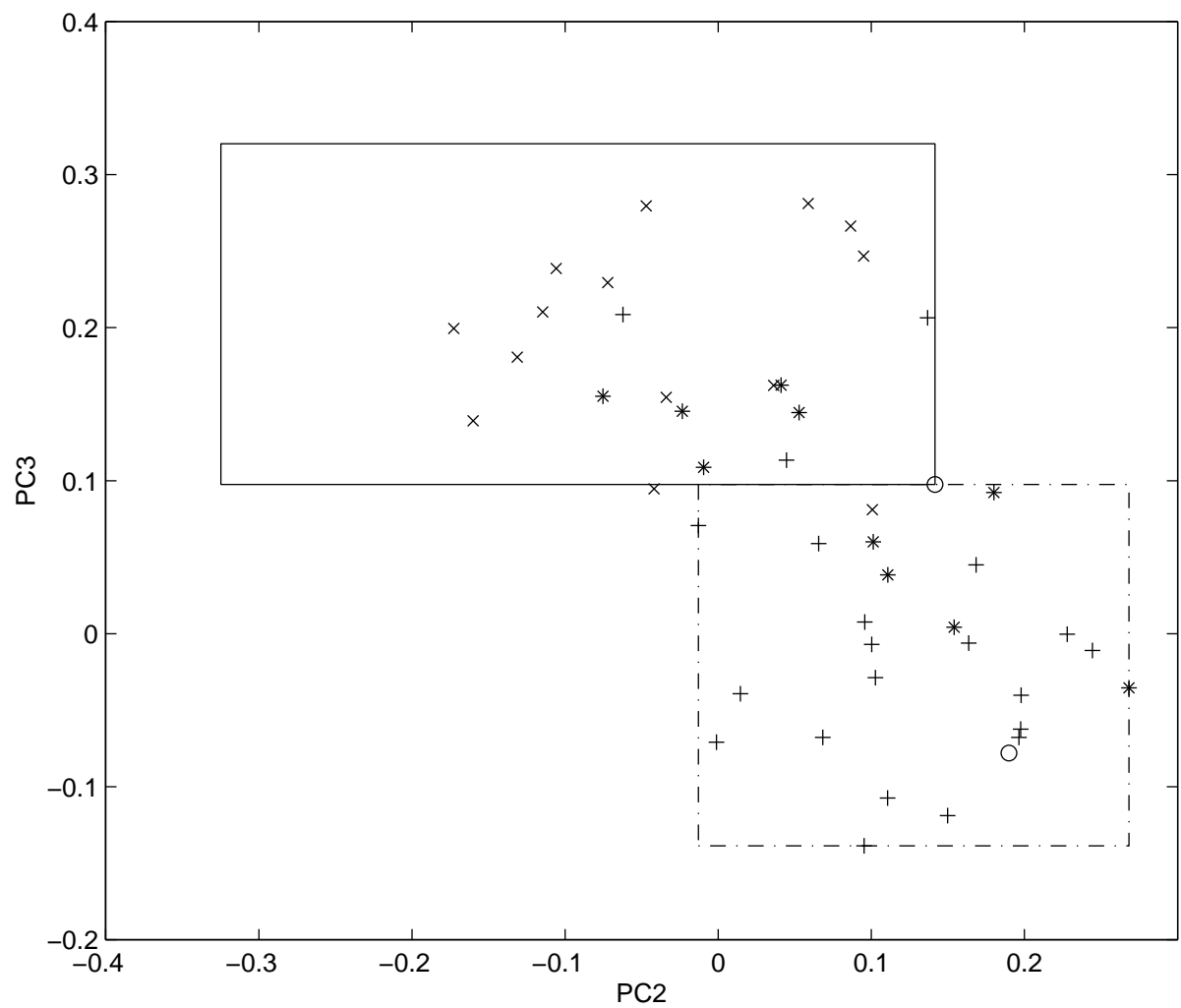
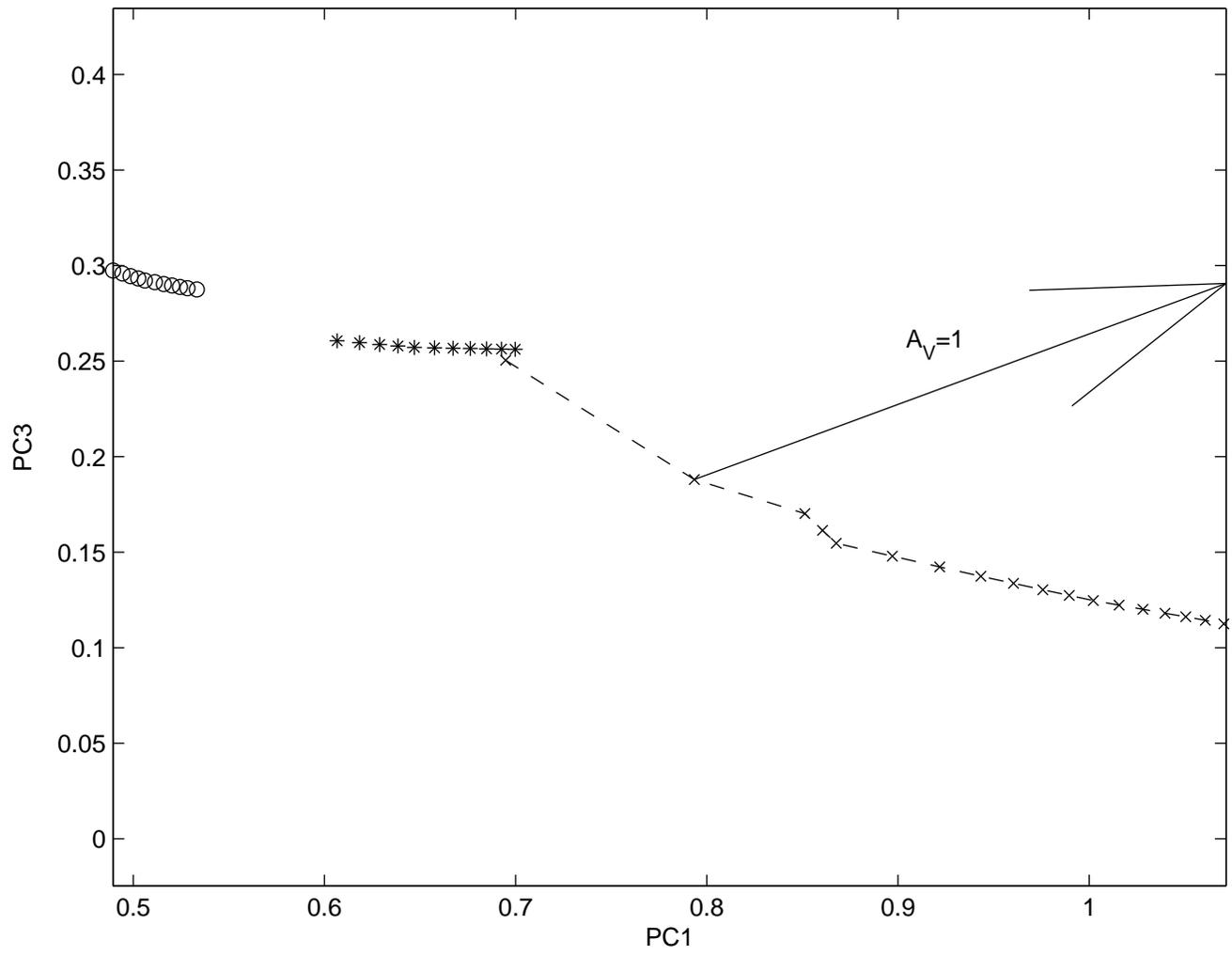
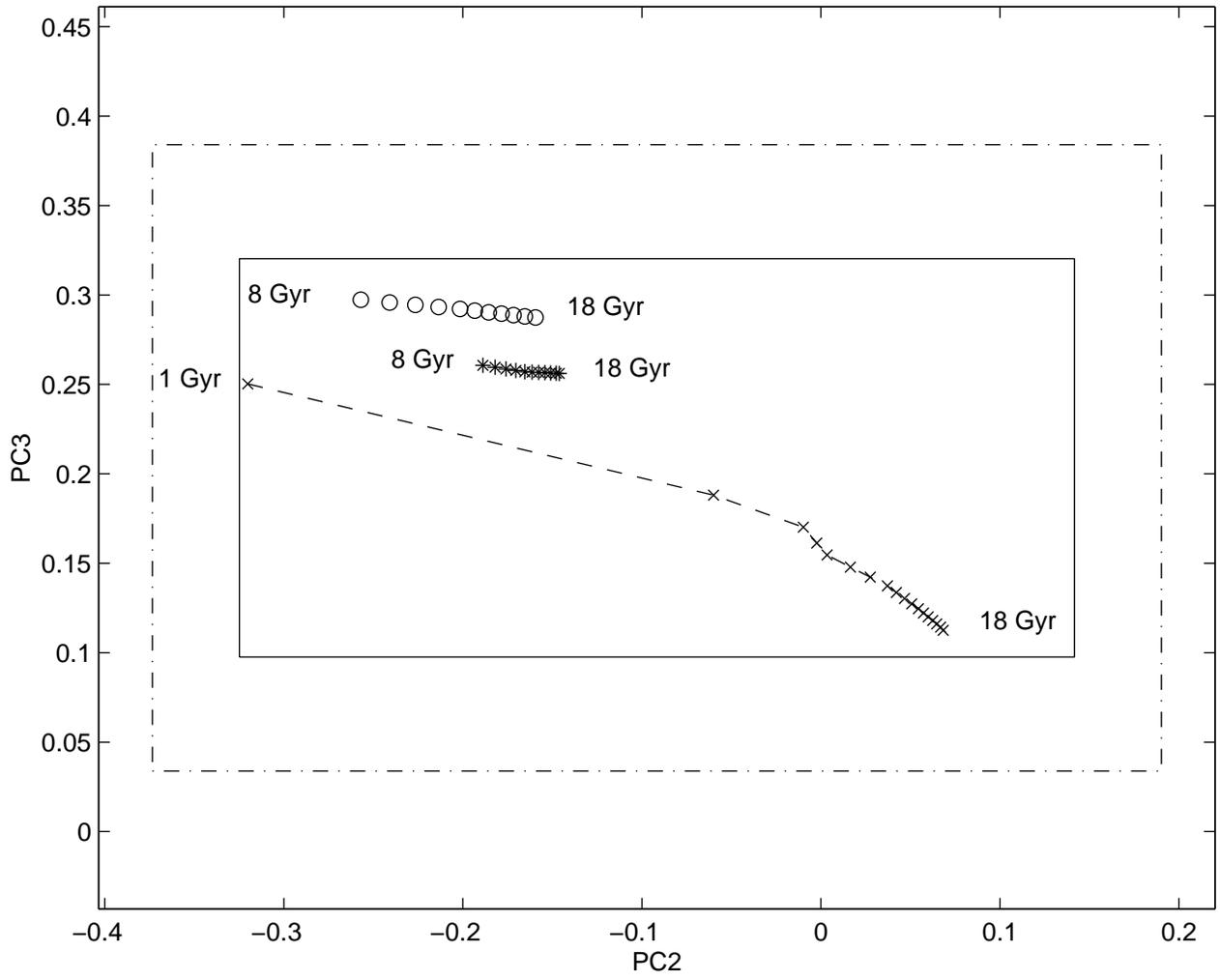
\end{document}